\documentclass[numberedappendix]{emulateapj}

\usepackage{color}

\newcommand{\SDSS}{SDSS--II}

\newcommand{\snana}{{\tt SNANA}}

\newcommand{\unc}{uncertainty}
\newcommand{\uncs}{uncertainties}

\newcommand{\mlcs}{{\sc mlcs2k2}}
\newcommand{\SALTII}{{\sc saltii}}

\newcommand{\zmlcs}{{\sc mlcs2k2+z}}
\newcommand{\zSALTII}{{\sc saltii+z}}
\newcommand{\zLCFIT}{{\sc lcfit+z}}

\newcommand{\LCDM}{\Lambda{\rm CDM}}

\newcommand{\PROBCHI}{P_{\chi^2}}
\newcommand{\CHISQ}{\chi^2}
\newcommand{\SIGMAi}{\sigma_i}
\newcommand{\XFIT}{\vec{x}_{5}}

\newcommand{\Z}{z}
\newcommand{\Zphotone}{\Z_{\rm phot}^1}
\newcommand{\Zphot}{\Z_{\rm phot}}
\newcommand{\Zspec}{\Z_{\rm spec}}

\newcommand{\Zgen}{\Z_{\rm gen}}
\newcommand{\ZSN}{\Z_{\rm SN}}
\newcommand{\Zgal}{\Z_{\rm gal}}

\newcommand{\DZSYM}{\Delta_{\Z}}
\newcommand{\DZDEF}{(\Zphot - \Zspec)/(1+\Zspec)}
\newcommand{\zsig}{\sigma_{\DZSYM}}
\newcommand{\zrms}{{\rm RMS}_{\DZSYM}}

\newcommand{\zresSNLS}{\sigma_{\Delta z/(1+z)}}

\newcommand{\NTOTASTIER}{55}
\newcommand{\NLOZASTIER}{13}
\newcommand{\NHIZASTIER}{42}

\newcommand{\zminf}{\Z_{{\rm min,f}}}
\newcommand{\zmaxf}{\Z_{{\rm max,f}}}

\newcommand{\zmini}{\Z_{{\rm min},i}}
\newcommand{\zmaxg}{\Z_{{\rm max},g}}

\newcommand{\Zmax}{Z_{\rm max}}
\newcommand{\cutZphotSYM}{\Delta {\Z}_{\rm phot}^{\rm cut}}
\newcommand{\cutphotodzSNLS}{0.04}
\newcommand{\cutphotodzSDSS}{0.03}

\newcommand{\ZBINSCAN}{0.04}
\newcommand{\CBINSCAN}{0.2}

\newcommand{\lamf}{\bar{\lambda}_{\rm f}}
\newcommand{\lammin}{\lambda_{\rm min}}
\newcommand{\lammax}{\lambda_{\rm max}}

\newcommand{\Trest}{T_{\rm rest}}
\newcommand{\Tobs}{T_{\rm obs}}

\newcommand{\photoz}{photo-$z$}

\newcommand{\specy}{spectroscopically}
\newcommand{\spec}{spectroscopic}

\newcommand{\eff}{efficiency}

\newcommand{\PEAKMAGSIGMA}{\sigma_{\rm f}}

\newcommand{\OM}{\Omega_{\rm M}}
\newcommand{\OL}{\Omega_{\Lambda}}

\newcommand{\mufit}{{\mu}_{\rm fit}}
\newcommand{\muref}{{\mu}_{\rm ref}}
\newcommand{\mugen}{{\mu}_{\rm gen}}
\newcommand{\muini}{{\mu}_{\rm ini}}
\newcommand{\murms}{{\rm RMS}_{\mu}}

\newcommand{\Ngrid}{N_{\rm grid}}
\newcommand{\marg}{marginalization}

\newcommand{\wwwSNANA}{\tt http://www.sdss.org/supernova/SNANA.html}
\newcommand{\wwwMINUIT}{\tt http://wwwasdoc.web.cern.ch/wwwasdoc/minuit/minmain.html}

\newcommand{\wwwPanSTARRS}{\tt http://pan-starrs.ifa.hawaii.edu/public}

\newcommand{\NCUTSLSSTDEEP}{1900}
\newcommand{\NCUTSLSSTMAIN}{5\times 10^4}
\newcommand{\minuit}{{\sc minuit}}
\newcommand{\wwwSDSS}{\tt http://www.sdss.org/}

\def\zsigSIMCOHSDSSMLCS{  0.020}   
\def\zsigeSIMCOHSDSSMLCS{  0.001}   
\def\zsigSIMCOHSDSSSALT{  0.017}   
\def\zsigeSIMCOHSDSSSALT{  0.001}   
\def\zsigSIMCOHSNLSMLCS{  0.021}   
\def\zsigeSIMCOHSNLSMLCS{  0.001}   
\def\zsigSIMCOHSNLSSALT{  0.014}   
\def\zsigeSIMCOHSNLSSALT{  0.001}   
 
\def\NDATASDSSMLCS{   296}   
\def\zsigDATASDSSMLCS{  0.031}   
\def\zsigeDATASDSSMLCS{  0.003}   
 
\def\zsigSIMSDSSMLCS{  0.030}   
\def\zsigeSIMSDSSMLCS{  0.001}   
 
\def\zsigDATASDSSSALT{  0.027}   
\def\zsigeDATASDSSSALT{  0.002}   
 
\def\zsigSIMSDSSSALT{  0.030}   
\def\zsigeSIMSDSSSALT{  0.001}   
 
\def\NDATASNLSMLCS{    37}   
\def\zrmsDATASNLSMLCS{  0.045}   
\def\zsigDATASNLSMLCS{  0.051}   
\def\zsigeDATASNLSMLCS{  0.020}   
 
\def\zsigSIMSNLSMLCS{  0.027}   
\def\zsigeSIMSNLSMLCS{  0.002}   
 
\def\NDATASNLSSALT{    37}   
\def\zrmsDATASNLSSALT{  0.040}   
\def\zsigDATASNLSSALT{  0.032}   
\def\zsigeDATASNLSSALT{  0.006}   
 
\def\zsigSIMSNLSSALT{  0.028}   
\def\zsigeSIMSNLSSALT{  0.002}   

\def\zbiasUNIT{\times 10^{-4}}
 
\def\zrmsSIMTESTA{  0.006}
\def\zbiasSIMTESTA{ ( -1.7 \pm  2.1) }

\def\zrmsSIMTESTB{  0.009}
\def\zbiasSIMTESTB{ (  0.5 \pm  2.9) }

\def\zrmsSIMTESTC{  0.015}

\begin{document}

\title{
Photometric Estimates of Redshifts and Distance Moduli for 
Type Ia Supernovae
}

\submitted{Accepted by ApJ}
\email{kessler@kicp.uchicago.edu}


\newcommand{\NUMUCASTRO}{1}   
\newcommand{\NUMKICP}{2}      
\newcommand{\NUMWAYNE}{3}     
\newcommand{\NUMRUTGERS}{4}   
\newcommand{\NUMFNAL}{5}      
\newcommand{\NUMIRISH}{6}    
\newcommand{\NUMPORT}{7}     
\newcommand{\NUMANL}{8}        
\newcommand{\NUMUPENN}{9}      
\newcommand{\NUMPENNSTATE}{10}  
\newcommand{\NUMCAPEMATH}{11}  
\newcommand{\NUMSAAO}{12}      
\newcommand{\NUMAPO}{13}       
\newcommand{\NUMSTOCK}{14}       
\newcommand{\NUMKLEIN}{15}       
\newcommand{\NUMLCO}{16}       
\newcommand{\NUMBOHR}{17}      


\author{
Richard~Kessler,\altaffilmark{\NUMUCASTRO,\NUMKICP}
David~Cinabro,\altaffilmark{\NUMWAYNE}
Bruce~Bassett,\altaffilmark{\NUMCAPEMATH,\NUMSAAO}
Benjamin~Dilday,\altaffilmark{\NUMRUTGERS}
Joshua~A.~Frieman,\altaffilmark{\NUMUCASTRO,\NUMKICP,\NUMFNAL}
Peter~M.~Garnavich,\altaffilmark{\NUMIRISH}
Saurabh~Jha,\altaffilmark{\NUMRUTGERS}
John~Marriner,\altaffilmark{\NUMFNAL}
Robert~C.~Nichol,\altaffilmark{\NUMPORT}
Masao~Sako,\altaffilmark{\NUMUPENN}
Mathew~Smith,\altaffilmark{\NUMCAPEMATH}
Joseph~P.~Bernstein,\altaffilmark{\NUMANL}
Dmitry~Bizyaev,\altaffilmark{\NUMAPO}
Ariel~Goobar,\altaffilmark{\NUMSTOCK,\NUMKLEIN}
Stephen~Kuhlmann,\altaffilmark{\NUMANL}
Donald~P.~Schneider,\altaffilmark{\NUMPENNSTATE}
Maximilian~Stritzinger\altaffilmark{\NUMLCO,\NUMBOHR}
} 


\altaffiltext{\NUMUCASTRO}{
  Department of Astronomy and Astrophysics,
   The University of Chicago, 5640 South Ellis Avenue, Chicago, IL 60637
}

\altaffiltext{\NUMKICP}{
  Kavli Institute for Cosmological Physics, 
   The University of Chicago, 5640 South Ellis Avenue Chicago, IL 60637
}

\altaffiltext{\NUMWAYNE}{
Department of Physics and Astronomy, 
Wayne State University, Detroit, MI 48202
}

\altaffiltext{\NUMRUTGERS}{
Department of Physics and Astronomy, 
Rutgers University, 136 Frelinghuysen Road, Piscataway, NJ 08854
}

\altaffiltext{\NUMFNAL}{
Center for Particle Astrophysics, 
  Fermi National Accelerator Laboratory, P.O. Box 500, Batavia, IL 60510
}

\altaffiltext{\NUMIRISH}{
  University of Notre Dame, 225 Nieuwland Science, Notre Dame, IN 46556-5670
}

\altaffiltext{\NUMPORT}{
  Institute of Cosmology and Gravitation, Mercantile House,
   Hampshire Terrace, University of Portsmouth, Portsmouth PO1 2EG, UK
}

\altaffiltext{\NUMUPENN}{
Department of Physics and Astronomy,
University of Pennsylvania, 203 South 33rd Street, Philadelphia, PA  19104
}

\altaffiltext{\NUMANL}{
Argonne National Laboratory, 9700 S. Cass Avenue, Lemont, IL 60437
}

\altaffiltext{\NUMPENNSTATE}{
  Department of Astronomy and Astrophysics,
   The Pennsylvania State University,
   525 Davey Laboratory, University Park, PA 16802.
}

\altaffiltext{\NUMCAPEMATH}{
Department of Mathematics and Applied Mathematics,
University of Cape Town, Rondebosch 7701, South Africa
}

\altaffiltext{\NUMSAAO}{
  South African Astronomical Observatory,
   P.O. Box 9, Observatory 7935, South Africa.
}

\altaffiltext{\NUMAPO}{
  Apache Point Observatory, P.O. Box 59, Sunspot, NM 88349.
}

\altaffiltext{\NUMSTOCK}{
Department of Physics, Stockholm University, Albanova University 
Center, SE--106 91 Stockholm, Sweden
}

\altaffiltext{\NUMKLEIN}{
The Oskar Klein Centre for Cosmoparticle Physics, Department of Physics,
AlbaNova, Stockholm University, SE-106 91 Stockholm, Sweden
}

\altaffiltext{\NUMLCO}{Las Campanas Observatory, Carnegie Observatories,
   Casilla 601, La Serena, Chile 
}

\altaffiltext{\NUMBOHR}{Dark Cosmology Centre, Niels Bohr Institute, University
of Copenhagen, Juliane Maries Vej 30, 2100 Copenhagen \O, Denmark 
}


\begin{abstract}
Large planned photometric surveys will discover hundreds of thousands 
of supernovae (SNe), outstripping the resources available for \spec\
follow-up and necessitating the development of purely photometric 
methods to exploit these events for cosmological study.
We present a light curve fitting technique for 
type Ia supernova (SN~Ia) photometric 
redshift ({\photoz}) estimation in which the redshift is determined 
simultaneously with the other fit parameters.
We implement this ``{\zLCFIT}'' technique within the 
frameworks of the \mlcs\ and \SALTII\ light curve fit methods
and determine the precision on the redshift and distance modulus.
This method is applied to
a \specy\ confirmed sample of
$\NDATASDSSMLCS$ SNe~Ia from the 
Sloan Digital Sky Survey-II  ({\SDSS}) SN Survey and 
$\NDATASNLSMLCS$ publicly available SNe Ia from the 
Supernova Legacy Survey (SNLS). 
We have also applied the method to a large suite of realistic
simulated light curves for existing and planned surveys, including
the SDSS, SNLS, and Large Synoptic Survey Telescope (LSST).
When intrinsic SN color fluctuations are included, 
the \photoz\ precision for the 
simulation is consistent with that in the data.
Finally, we compare the \zLCFIT\ \photoz\ precision with previous 
results using color-based SN \photoz\ estimates. 
\end{abstract}


\section{Introduction}
\label{sec:intro}

To investigate the expansion history of the universe, 
increasingly large samples of high-quality type Ia supernova
(SN~Ia) light curves
are being used to measure luminosity distances as a function of
redshift (the SN Ia Hubble diagram).  
Expected SN~Ia samples will be in the thousands for the 
Dark Energy Survey (DES: \citet{DES-moriond2009}) 
and in the hundreds of thousands for the surveys to be 
carried out by the 
Panoramic Survey Telescope and Rapid Response System
(Pan-STARRS)\footnote{\wwwPanSTARRS}
and by the Large Synoptic Survey Telescope
(LSST: \citet{Ivezic_08,LSSTSB09}).
For the latter two, only a small fraction of the SNe will have 
\specy\ determined redshifts from SN or host-galaxy spectra 
for the foreseeable future.
To make use of these large SN samples, the redshifts will have
to be determined photometrically using both the SN light curves
and the host-galaxy photometric observables.

Methods for estimating galaxy photo-$z$'s have been developed over 
many years (for a review, see, e.g., \citet{Abdalla08}).
They generally fall into two categories: (1) an empirical 
approach, in which one translates observed colors, magnitudes, or
other photometric observables  
into a redshift estimate, training the algorithm on a subset of galaxies 
with spectroscopic redshifts; and (2) template fitting, in which observed 
colors are matched to redshifted template galaxy 
spectral energy distributions until the best 
match for galaxy redshift and type is obtained. The development of 
photo-$z$ methods using SN data is more recent. Some 
have followed and adapted the empirical approach to galaxy 
photo-$z$ estimation, e.g., using observed SN colors near the epoch of 
peak brightness to estimate the redshift
\citep{YWang2007_SNLS,YWang2007_SIM} 
or early-epoch colors to select SNe in particular redshift regions 
\citep{Dahlen2002}.

In this paper, we present and describe a method of SN~Ia \photoz\ 
estimation that is analogous to the template fitting method 
for galaxy {\photoz}'s. In models used to fit SN~Ia light curves, 
one typically uses a \specy\ determined redshift, 
and the fit parameters are usually taken to be the epoch of 
peak brightness, the light curve shape or stretch, 
a color or dust extinction estimate, and the distance modulus. 
In our approach, we extend the usual methods of fitting light curves 
to include the redshift as a fifth fit parameter. 
We apply this ``{\zLCFIT}'' method to determine {\photoz}'s 
for the \specy\ confirmed \SDSS\ \citep{Frieman07,K09} and 
SNLS (\citet{Astier06}; hereafter A06) SNe,
and compare the resulting photo-$z$ precision to that obtained 
from simulated samples. Once we have verified the reliability 
of the simulations by comparison with the data, 
we apply the \zLCFIT\ method to simulated  LSST SN observations.
In all cases, we use both \mlcs\ \citep{JRK07,K09} and 
\SALTII\ \citep{Guy07} light curve fitting models.

Variants of the light curve fit approach to SN  {\photoz}'s
have been used before, by both the SNLS \citep{Sullivan2006} 
and SDSS \citep{Sako08} surveys, to select SN Ia candidates for
\spec\ follow-up after only a few photometric epochs. 
\citet{KimMiquel07} used the {\SALTII} model and a Fisher matrix
analysis to estimate \uncs\ on photometric redshifts and distances.
Using a technique similar to our {\zLCFIT} method,
\citet{Gong09} studied LSST simulations,
focusing mainly on contamination from non-Ia SNe
and the resulting precision on cosmological parameters.
In contrast, we focus here on the 
precision and bias for the \photoz\ and distance modulus.
We also use more realistic simulations based on the LSST cadence 
for both the deep and wide surveys and illustrate some differences
between these two components of the LSST survey.
Our estimates of non-Ia contamination and cosmological precision
will be presented in a future work. Recently, 
\citet{SNLS_PHOTOZ09} (hereafter PD09) employed the 
light curve fit photo-$z$ technique within the \SALTII\ framework.
For nearly 300 SNe~Ia from the SNLS, they evaluated both the 
photo-$z$ precision and the fraction of catastrophic redshift outliers. 
The PD09 SN sample is by far the largest to date used to 
study SN \photoz\ methods; our \SDSS\ sample, 
at lower redshifts, is of comparable size.

The plan of this paper is as follows. We introduce the 
\specy\ confirmed SN data samples (\SDSS\ and SNLS)
in \S\ref{sec:data}. In \S\ref{sec:sim}, we describe the 
simulation, and we present the \zLCFIT\ method in detail in 
\S\ref{sec:zLCFIT_method}. 
The \photoz\ precision and fit-parameter correlations
for the data samples, along with the corresponding 
results from simulated samples, 
are presented in \S\ref{sec:results_data}.
We use the simulation to make predictions for the 
LSST survey in \S\ref{sec:lsst}.
In \S\ref{sec:compare} we make direct comparisons with the 
color-based \photoz\ method presented in previous works.

As described in \citet{SNANA09},
all light curve fitting and simulation software is
publicly available in the \snana\ package.\footnote{\wwwSNANA}

 \section{The \SDSS\ and SNLS Data Samples}
 \label{sec:data}

To test the \zLCFIT\ method, we use the full
three-season sample from the \SDSS\ Supernova (SN) Survey 
\citep{Frieman07},
and the publicly available sample from the first season 
of the SNLS (A06).
Below we give a brief description of these samples.

The \SDSS\ SN Survey used the SDSS camera \citep{Gunn_98} 
on the SDSS 2.5~m telescope \citep{SDSS_telescope} at 
Apache Point Observatory to search for SNe in the Fall seasons
(September 1 through November 30) of 2005--2007.
This survey scanned a region (designated stripe~82) 
centered on the celestial equator in the 
Southern Galactic hemisphere that is 
2.5$^{\circ}$ wide and runs between right ascensions of
20$^{\rm hr}$ and 4$^{\rm hr}$, 
covering a total area of $300~{\rm deg}^2$ with a typical cadence 
of every four nights per region. 
Images were obtained in five broad passbands,
$ugriz$ \citep{Fukugita_96}, with 55~s exposures and 
processed through the PHOTO photometric pipeline \citep{Lupton_01}. 
Within 24~hours of collecting the data, the images were 
searched for SN candidates that were selected for \spec\ follow-up 
observations in a program involving about a dozen telescopes.
The \SDSS\ SN Survey discovered and \specy\ confirmed
a total of $\sim 500$ SNe~Ia. 
A larger sample of photometrically identified but \specy\ 
unobserved SNe~Ia was also compiled, and host-galaxy redshifts 
for several hundred of these photometric candidates have been 
obtained to date. The SDSS-III Survey \citep{BOSS_09}, 
as a small part of its early program, is in the process 
of measuring host-galaxy redshifts for more than 1000 of these 
photometrically identified SNe~Ia. 
The telescope aperture, focal plane, and exposure time of the 
SDSS system \citep{York_00} were ideal for discovering SNe in 
the previously underexplored redshift range $0.1 < z < 0.3$.
Details of the SDSS-II SN Survey are given in \citet{Frieman07,Sako08},
the procedures for \spec\ identification and redshift determinations
are described in \citet{Zheng08}, and the SN photometry is described in 
\citet{Holtz08}. A condensed summary of the \SDSS\ survey, SN typing, 
redshift determination, photometry, and calibration 
can be found in \citet{K09}.

The SNLS was a five-year survey covering $4~{\rm deg}^2$ 
using the MegaCam imager on the 3.6~m 
Canada--France--Hawaii Telescope (CFHT). 
Images were taken in four bands similar to those used
by the SDSS: $g_M, r_M, i_M, z_M$, where 
the subscript $M$ denotes the MegaCam system.
The SNLS exposures were $\sim 1$~hr in order
to discover SNe at redshifts up to $z\sim 1$.
The SNLS images were processed in a fashion similar to the 
\SDSS\ so that \spec\ observations could be used to confirm the 
identities and determine the redshifts of the SN candidates.
We use the publicly available sample from their first year
of operations that ended July 15, 2004.
Detailed information about the SNLS can be found in A06
and references therein.

Since high-quality light curves are needed for the
\zLCFIT\ method, we apply the following selection requirements 
to the photometric data for inclusion in our analysis samples:
(1) spectroscopic confirmation of type Ia,
(2) a measurement with $\Trest < -3$ days,
(3) a measurement with $\Trest > +10$ days,
(4) measurements in at least three observer-frame filters 
have signal-to-noise ratio (S/N) greater than 8, and
(5) the probability corresponding to the fit-$\chi^2/N_{\rm dof}$
(\S\ref{sec:zLCFIT_method}) is $\PROBCHI > 0.02$.
Here $\Trest$ is the epoch in the SN rest frame relative to peak
brightness in the $B$ band, and we note that it depends on the 
fitted \photoz\ value. For the \SDSS, we use only the $gri$ passbands.
The number of SNe~Ia satisfying these selection requirements
is nearly 300 and 40 for the \SDSS\ and SNLS, respectively;
the exact numbers of SNe depend on the fitting model 
({\mlcs} or {\SALTII})
and will be given in \S\ref{sec:results_data}.
The selection criteria above are not based on optimizations,
but are instead motivated by the strong correlation between
redshift and color. Requiring two colors and at least one measurement 
in each of three passbands with ${\rm S/N}>8$ explicitly
ensures good color measurements. 
The $\Trest$ requirements ($<-3$ and $>+10$~days) ensure a good 
determination of the time of peak brightness ($t_0$);
since SNe become redder after peak light, a mismeasurement of 
$t_0$ translates directly into a mismeasurement of color, 
and hence redshift.

Here we provide some additional motivation for 
the above requirements.
Relaxing the S/N requirement (no. 4 above) to ${\rm S/N}>5$ 
results in about 10\% more SNe~Ia in the \SDSS\ sample, 
and a 20\% degradation in the \photoz\ precision. 
Making a more restrictive cut of
${\rm S/N}>10$ results in a 20\% loss of events and a
negligible improvement in the precision. 
We have therefore chosen ${\rm S/N}>8$ as a reasonable
compromise between sample statistics and precision.
To motivate the sampling requirements,
we have applied the \zLCFIT\ fitting method 
(\S\ref{sec:zLCFIT_method}) to the \SDSS\ sample in which 
all measurements prior to peak brightness have been rejected;
the resulting precision and bias on the fitted $t_0$ and \photoz\ 
are significantly degraded.
The sampling requirements above (nos. 2 and 3) are therefore 
designed to ensure a good determination of $t_0$.

\section{Simulations}
\label{sec:sim}

We use the \snana\ simulation code to generate realistic 
SN~Ia light curves that can be analyzed in exactly the 
same manner as the data. 
The simulation is used to compare with the data,
to compare with previous \photoz\ studies based on simulations,
and to make predictions for  LSST.
All SN simulations are based on a standard
$\LCDM$ cosmology ($w=-1$, $\OM=0.3$, $\OL=0.7$) 
and they are generated and fit using the same light curve model:
\mlcs\ or {\SALTII}. 
This strategy explicitly assumes that the light curve model 
is correct, and will therefore yield the most optimistic
results. As discussed below, the models are adjusted to 
account for the anomalous Hubble scatter. However, these
models have not been adjusted to account for the 
discrepancies in the ultraviolet region 
(\citet{K09}: hereafter K09)
Details of the simulation are described in \citet{SNANA09} 
and in \S 6 of K09; here we give a brief overview.

For \mlcs, rest-frame model magnitudes are generated
from light curve templates and then dimmed by  
host-galaxy dust extinction. 
Using SN~Ia spectral templates from \citet{Hsiao07},
$K$-corrections \citep{Nugent2002}
are used to transform the rest-frame ($UBVRI$) model  
magnitudes into observer-frame magnitudes.
For \SALTII, the model is based on
a time-sequence of rest-frame spectra, 
and observer-frame magnitudes are computed
by convolution with the appropriate filter-response curves.

Since light curve models are defined over a specific wavelength 
range in the rest frame, one usually checks the rest-frame
wavelength $\lamf/(1+\Z)$, where $\lamf$ is the mean wavelength 
of the observer-frame filter and $\Z$ is the redshift. 
If the rest-frame wavelength is outside the valid range of the model,
then the corresponding filter is typically ignored. 
For \photoz\ applications, this method of ignoring filters 
clearly cannot be used, since the redshift is not known ahead of time. 
To be realistic, we should not use this filter-ignoring procedure 
in the simulations either. 
Therefore, our simulations use wavelength-extended models that generate 
fluxes for filters with $\lamf/(1+\Z)$ beyond the nominally defined
wavelength range. For both \mlcs\ and \SALTII, the rest-frame
wavelengths are extended down to 2500~\AA.
For \SALTII, the 7000~\AA\ upper limit has been raised to 8700~\AA.
We note that the simulation and fitter use the same
wavelength-extended model, and therefore these tests do not probe
potential problems if the extended part of the model is wrong.

We have implemented two models of intrinsic SN magnitude variations
that produce ``anomalous'' scatter in the Hubble diagram.
The source of anomalous scatter is unknown, and it is not clear
if the scatter can be reduced with an improved light curve model, 
or if this scatter is due to some random physical process such 
as brightness variations as a function of viewing angle in an 
asymmetric explosion.
For simulations with $10^4$ times the nominal exposure time
and $\Z < 0.5$, i.e., for which photon noise is negligible, 
we define the anomalous scatter to be the rms ($\murms$) 
of the difference between the fitted and generated 
distance modulus ($\mufit - \mugen$) from the four-parameter 
light curve fit using \specy\ determined redshifts.
The default model, called ``color-smearing,'' 
introduces an independent magnitude fluctuation in each passband,
and the fluctuation is the same for all epochs within each passband.
A random number $r_{f}$ from a unit-variance Gaussian distribution 
is chosen for each passband $f$, and  
a magnitude fluctuation  
$\delta m_{f} = r_{f} \PEAKMAGSIGMA$
is added to the generated magnitude at all epochs.
As described in \S\ref{sec:results_data}, a scatter of
$\PEAKMAGSIGMA=0.1$~mag is needed in order for the
simulated \photoz\ precision to match that of the data.
The resulting Hubble scatter is consistent with that seen in
analyses of {\specy} confirmed data samples;
$\murms \simeq 0.16$ for the SNLS ($griz$) simulations, and 
$\murms \simeq 0.19$~mag for the \SDSS\ ($gri$) simulations.
The simulated SNLS scatter is slightly smaller because this
survey results in larger S/N values.
The second model of intrinsic variations is called 
``coherent luminosity smearing:''
a coherent random magnitude shift, 
drawn from a Gaussian with $\sigma_{\rm coh} \sim 0.15$~mag,
is added to the model magnitude for all epochs and passbands.
In the coherent smearing method 
the intrinsic model colors are not varied, and
the resulting anomalous scatter is $\murms = \sigma_{\rm coh}$.

Although both models of intrinsic magnitude variation 
result in the expected scatter in the Hubble diagram,
only the color-smearing model can generate
the observed \photoz\ precision in the \SDSS\ and SNLS
data samples (\S\ref{sec:results_data}). 
We have not investigated simulations in which both
models of intrinsic variation contribute,
nor have we investigated variations in the color parameter 
($R_V$ for {\mlcs} or $\beta$ for {\SALTII}) that could also
introduce anomalous scatter.

To simulate non-photometric conditions and varying 
time intervals between observations due to bad weather,
actual observing conditions are used for an existing survey, 
or an estimate of such conditions for a planned (future) survey.
For each simulated observation, the noise is determined from
the measured point spread function (PSF),\footnote{The PSF is
described by a double-Gaussian function for the \SDSS,
and by a single Gaussian for the other surveys.} 
zero point, CCD gain, and sky background.
Noise from the host-galaxy background is not included.
The simulated flux in CCD counts is based on a
mag-to-flux zero point and a random fluctuation drawn 
from the noise estimate.
For the \SDSS\ and SNLS surveys, 
a detailed treatment of the search \eff,
including \spec\ selection effects, is described in 
\S 6.2 of K09.

\begin{figure}[hb]
\centering
 \epsscale{1.1}
 \plotone{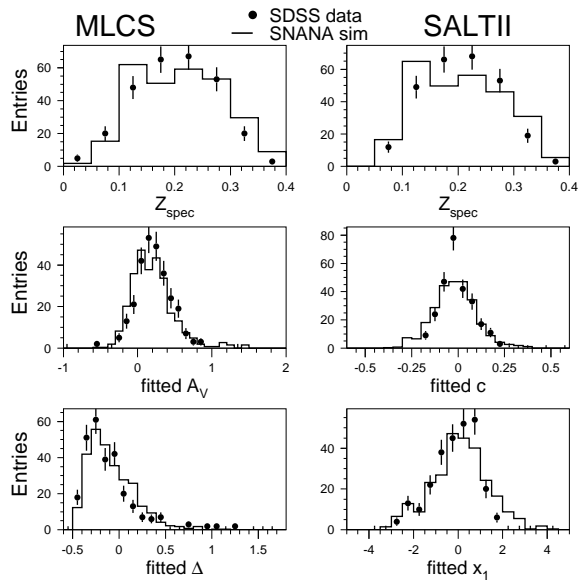}
  \caption{
	For the \SDSS\ SN Survey, comparison of data 
	distributions (dots) with those from the simulation (histograms).
	Fitted parameters for \mlcs\ ($z, A_V, \Delta$) are on the left, 
	and for \SALTII\ ($z, c, x_1$) on the right.	
	Fits were done with the redshift fixed to the true redshift.  
	The simulated histograms are scaled
	to have the same number of entries as the data.
      }
  \label{fig:ovdatasim_SDSS4par}
\end{figure}

\begin{figure}[hb]
\centering
 \epsscale{1.1}
 \plotone{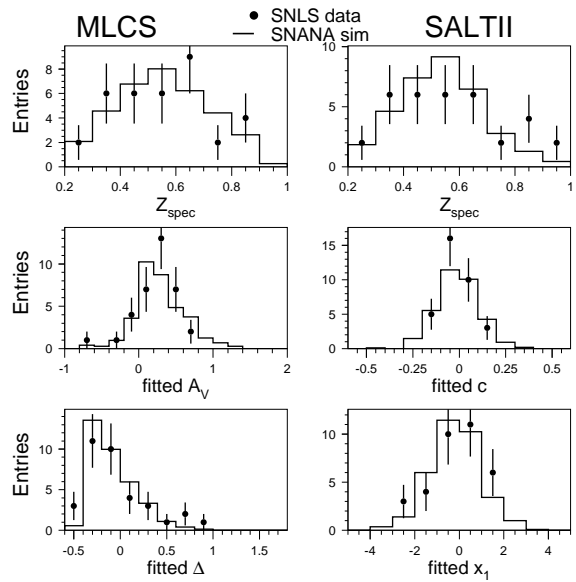}
  \caption{
	Same as Figure~\ref{fig:ovdatasim_SDSS4par}, 
	but for the SNLS survey, using the 
	data sample from A06.
      }
  \label{fig:ovdatasim_SNLS4par}
\end{figure}

The quality of the simulation is illustrated in 
Figures~\ref{fig:ovdatasim_SDSS4par} and \ref{fig:ovdatasim_SNLS4par} 
for the \SDSS\ and SNLS surveys, respectively,
using the SN selection requirements described in \S\ref{sec:data}.
The parameters from each sample have been determined 
using the conventional fitting method with a fixed 
\specy\ determined redshift.
Each figure shows data-simulation comparisons for the 
distributions of \spec\ redshift, color parameter 
($A_V$ for {\mlcs}, $c$ for {\SALTII}), 
and shape-luminosity parameter 
($\Delta$ for {\mlcs}, $x_1$ for {\SALTII}). 
The color and shape parameters are defined in \S\ref{sec:zLCFIT_method}.
There is good overall consistency between the measured and 
simulated distributions.

\section{The \zLCFIT\  Method}
\label{sec:zLCFIT_method}

\begin{figure}  
\centering
 \epsscale{1.0}
 \plotone{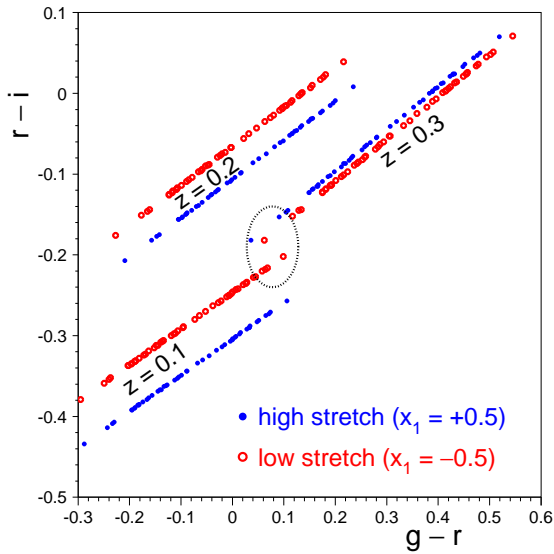}
  \caption{
	Observer-frame $r-i$ vs. $g-r$ at peak brightness 
	for SNe~Ia using the \SALTII\ model.
	Each set of color bands corresponds to the indicated redshift
	(0.1, 0.2, 0.3), and the variation within each band corresponds
	to variations in the intrinsic SN color 
	(i.e., the \SALTII\ $c$ parameter).
	The solid and open circles correspond to a high-stretch
	and  low-stretch SN, respectively.
	The dotted oval shows a degenerate region discussed
	in the text.
      }
  \label{fig:zcolor}
\end{figure}

The general principle behind the SN~Ia \photoz\ determination 
is illustrated in Fig.~\ref{fig:zcolor} using colors at the 
epoch of peak brightness.
Increasing the redshift causes uniform reddening at
all wavelengths, while intrinsic reddening (or extinction)
causes more reddening at bluer wavelengths.
A time-dependent light curve model (e.g., \mlcs\ or {\SALTII}) 
is used to account for the known color dependence on the
light curve shape (commonly known as the stretch) 
and to use all epochs in order to maximize the photostatics
in the \photoz\ measurement.
Since SNe become redder with increasing epoch, 
any error in the epoch of peak brightness results in the 
wrong template colors, and hence an increased error in the \photoz.

For real observations, the ideal color-color bands in
Fig.~\ref{fig:zcolor} are smeared by photon statistics
and by intrinsic SN color variations that are not described
by the light curve model.
With sufficient smearing, the colors of a very reddened SN at 
$z \simeq 0.1$ are degenerate with a blue SN at $z \simeq 0.3$
(see the region enclosed by dotted oval in Fig.~\ref{fig:zcolor})
and we indeed see this degeneracy in the \SDSS\ 
\photoz\ measurements. 
From data-simulation comparisons, we find that smearing from 
photon statistics does not fully describe the \photoz\  precision,
and we therefore propose that the modeling of intrinsic color 
variations is not adequate.
(\S \ref{sec:results_data}).

While host-galaxy photometric redshifts depend
on the determination of the 4000~{\AA} break,
a similar $\sim 2800$~{\AA} break in the SN spectrum
has little impact on the \photoz\ measurement because
this feature is either inaccessible at low redshifts,
or it is poorly measured at higher redshifts.
In current SN~Ia models, fluxes at these very blue 
wavelengths are either ignored, or they are heavily 
down-weighted relative to the optical bands.

The basic idea of the \zLCFIT\ method is to start with a 
light  curve fit model that has four free parameters when the 
redshift is fixed to an accurately measured value, 
and simply float the redshift as a fifth fitted parameter.
We refer to these methods as \zmlcs\ and \zSALTII,
where \mlcs\ and \SALTII\ refer to the conventional models
in which the redshift is fixed to a precisely measured value.
Compared to the color-based \photoz\ method, advantages of the 
\zLCFIT\ method include a natural framework for tracking correlations 
between redshift and distance modulus (\S\ref{subsec:fitcorr}), 
and using all of the light curve information 
(instead of just peak flux) so that in principle 
the intrinsic SN color variations can be accounted for.
The main advantage of the color-based method 
is that it works over a broader redshift range,
and there is no need to worry about which observer-frame 
filters map into a valid wavelength range in the rest frame.

\begin{figure}  
\centering
 \epsscale{1.0}
 \plottwo{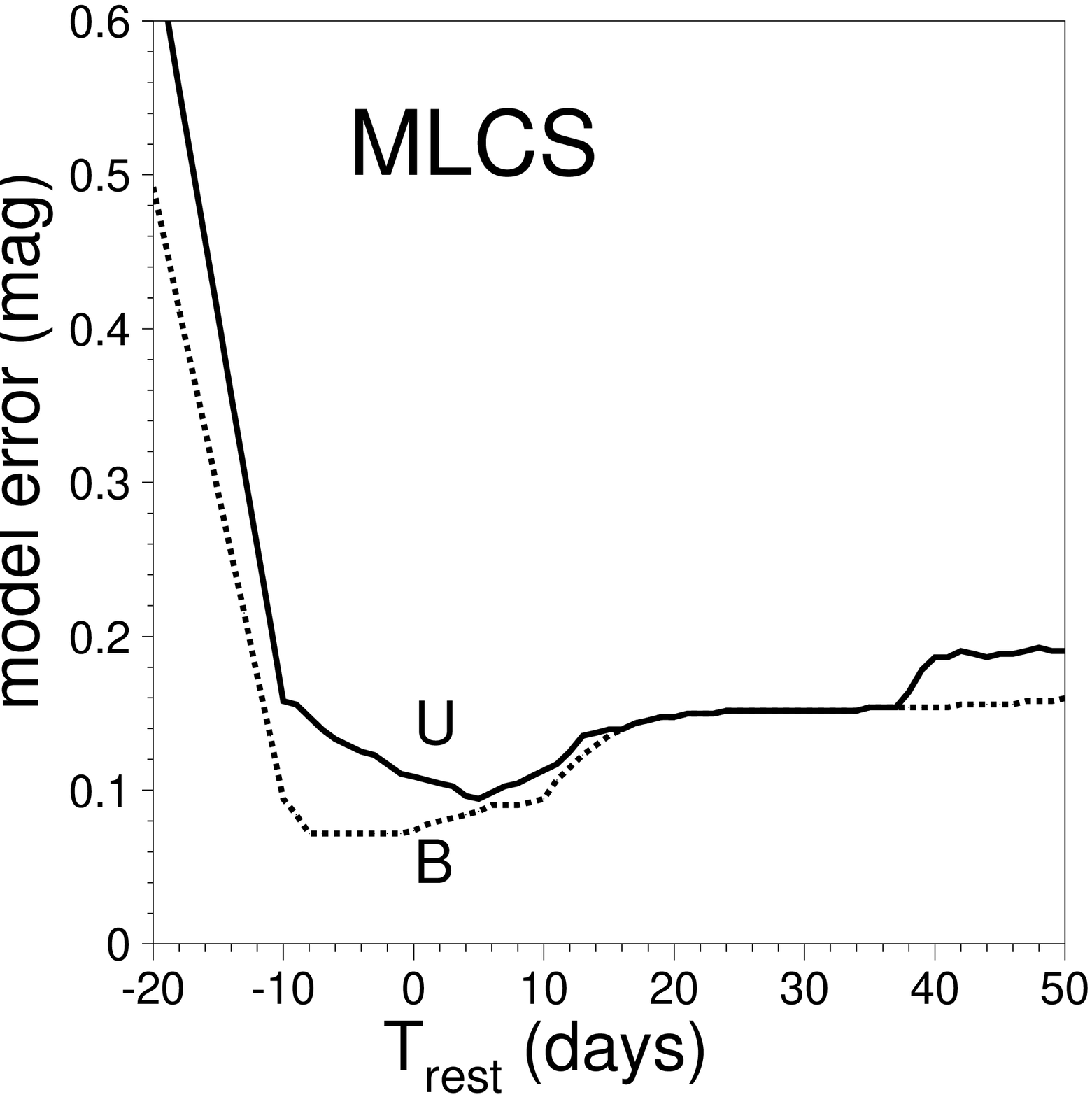}{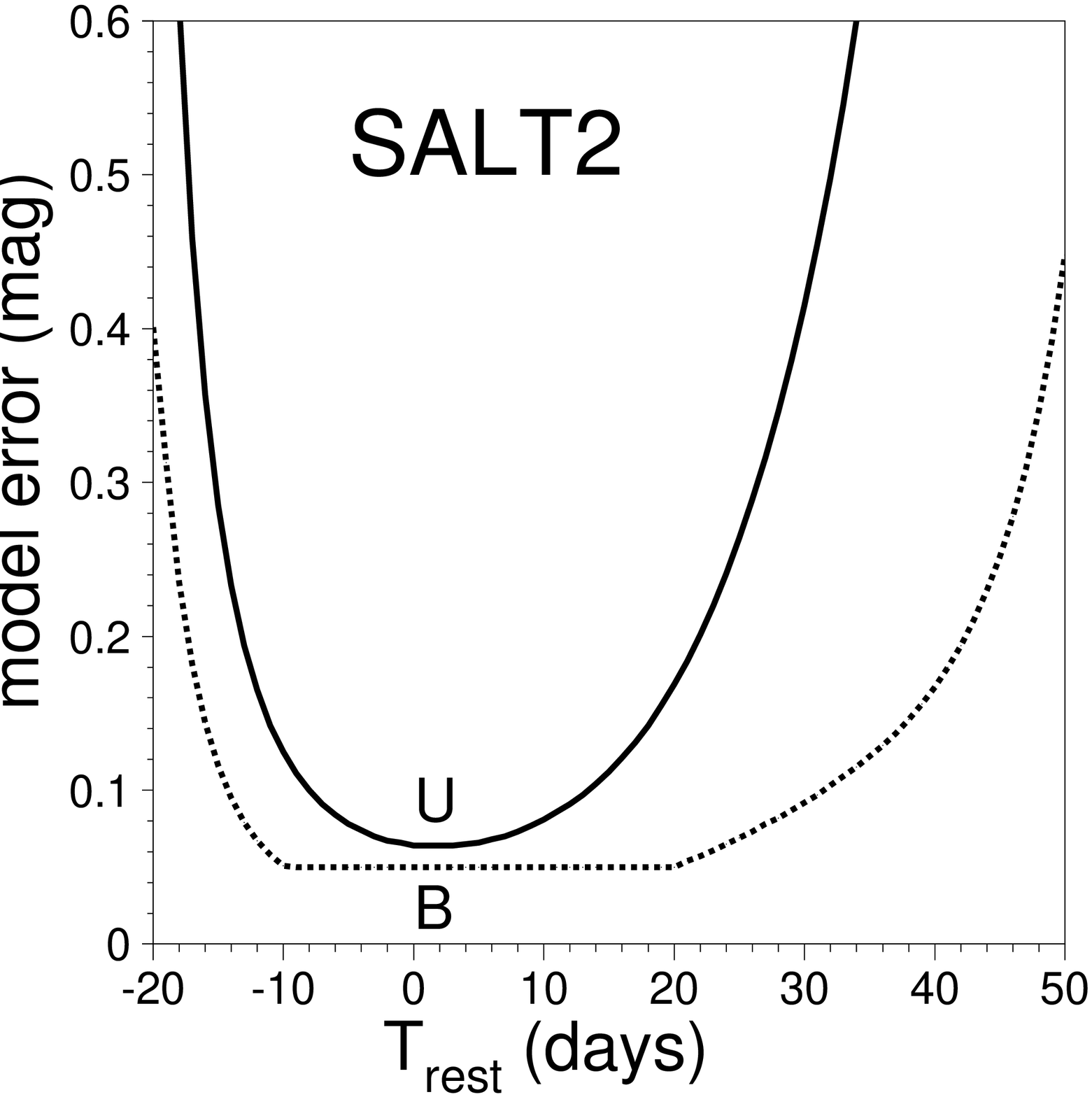}
  \caption{
	Model \unc\ vs. rest-frame epoch for the $U$ (solid)
	and $B$ (dashed) passbands.
      }
  \label{fig:modelerr}
\end{figure}

For \zmlcs, the five fitted parameters 
(which we denote with the vector $\XFIT$) are 
the time of maximum brightness in the (rest frame) $B$-band ($t_0$),
the shape-luminosity parameter ($\Delta$),
the host-galaxy extinction in the $V$ band ($A_V$),
the distance modulus ($\mu$),
and the redshift ($z_{\rm phot}$).
We use a flat $A_V$ prior and $R_V= A_V/E(B-V) = 2.2$.
For \zSALTII\ the five parameters are the 
time of maximum brightness in the rest frame $B$-band ($t_0$),
the shape-luminosity parameter ($x_1$),
the $B-V$ color ($c$),
the flux normalization ($x_0$),
and  the redshift ($z_{\rm phot}$).
The following light curve fit $\CHISQ$ is minimized using
{\minuit}\footnote{\wwwMINUIT}:
\begin{equation}
   \CHISQ = 
    \sum_i \left\{
      \frac{\left[F_i^{\rm data} - F_i^{\rm model}(\XFIT)\right]^2}
              {  \SIGMAi^2  }
       + 2\ln(\SIGMAi/\tilde{\SIGMAi})
        \right\}~,
  \label{eq:chi2}
\end{equation}
and the corresponding probability ($e^{-\CHISQ/2}$) is
used to marginalize as described in Appendix~\ref{app:marg}. Here, 
$F_i^{\rm data}$ is the SN flux of the $i$th observation,
$F_i^{\rm model}(\XFIT)$ is the predicted flux
using the five model parameters ($\XFIT$), and
$\SIGMAi^2 = \sigma_{i,{\rm stat}}^2 + \sigma_{i,{\rm model}}^2$
is the quadrature sum of the measured and model \uncs, respectively.
The index $i$ runs over all epochs and filters.
The second term in Eq.~\ref{eq:chi2} accounts for
model \uncs\ that depend on the rest-frame passband and epoch,
which in turn depend on the \photoz\ value.
The model \uncs\ are shown in Fig.~\ref{fig:modelerr}
for the rest-frame $U$ and $B$ passbands.
The reference \unc\ ($\tilde{\SIGMAi}$) from the first iteration 
of the fit is used in the denominator so that the second term is 
close to zero in the second iteration; 
although the $\tilde{\SIGMAi}$ do not affect the minimization,
these terms reduce the change in the calculated fit probability.
As explained below, the $\CHISQ$ is minimized twice in order to
include the appropriate filters and epochs. 
The minimized values and \uncs\ are then used to estimate
the integration ranges needed to obtain marginalized results.

In this study, we use the \mlcs\ and \SALTII\ light curve fitters
that have been implemented in the \snana\ package.
The main reasons for using the \snana\ code are 
(1) exactly the same light curve model is guaranteed to be used 
in both the light curve fits and in the generated simulations,
(2) there are several improvements to the \mlcs\ light curve fitter
as explained in K09,
(3) the \photoz\ implementation is identical for both models,
(4) the \snana\ fitter is significantly faster than the 
original fitting software.
To check the \SALTII\ implementation in \snana, 
we have repeated the light curve fits and cosmology analysis 
for the six sample combinations in K09 and find 
that the dark energy equation of state parameter $w$ is
always within a few hundredths of the value obtained 
with the original code;
these discrepancies are well below the statistical \uncs.

Although it is straightforward to include the redshift 
as a free parameter in the light curve fit,
there are subtle pre-fit issues related to 
the unknown redshift value:
(1) SN selection criteria that depend on knowing 
$\Trest$ = $\Tobs/(1+\Z)$,\footnote{$\Trest$ and $\Tobs$ are the 
rest-frame and observer-frame times in days 
since peak brightness in the $B$ band.
} 
such as requiring measurements with a 
minimum and maximum $\Trest$ value;
(2) as noted above, determining which observer-frame filters
(with mean wavelength $\lamf$) 
have $\lamf/(1+\Zphot)$ within the
valid wavelength range of the fitting model; 
(3) determining the valid rest-frame epoch range for the 
fitting model;
(4) as $\lamf/(1+\Zphot)$ maps into a different rest-frame filter 
    (for {\mlcs}) there is a discontinuous change in the model error,
    and therefore the $\chi^2$ is not a continuous function of 
    $\Zphot$; and
(5) determining robust initial fit-parameter values.
Our treatment of these issues is described in
Appendix~\ref{app:prefit}.


We end this section with a discussion of the processing time.
For the \SDSS\  light curves, 
which have nearly 50 measurements on average,
all of the minimization fit-iterations take $\sim 1$~s 
per SN using \minuit. 
The \marg\ (Appendix~\ref{app:marg})
takes close to half a minute per SN using 
an integration grid of 11 points per fit parameter, 
or a total of $11^5$ integration cells. 
However, the integration ranges usually need adjustment
after marginalizing, and therefore the marginalization 
typically runs twice, taking nearly a minute per SN.
The processing time scales linearly with the number
of measurements and as the fifth power of the number
of grid points per fit parameter.
For the integration grid above, the Monte Carlo Markov Chain
technique requires about the same amount of processing time.

Using the \snana\ implementation of \zmlcs\ and \zSALTII,
we find that \minuit\ gives adequate minimized values,
but that the \uncs\ are not reliable because of
subtle discontinuities in the $\chi^2$ derivative
with respect to the \photoz.
We must therefore marginalize in order to get
useful \uncs\ and covariances.
To illustrate the computational issue more clearly,
consider an LSST sample of 500,000 SNe~Ia.
The marginalization for all of the SNe in this sample 
requires a total of 1 CPU-year.
A factor of 100 is probably needed for code development
and systematic studies and another factor of several
for simulation studies. The total computing needs are 
therefore a few CPU centuries with today's processors,
assuming that $11^5$ integration cells give sufficient
accuracy. To use the minimization, which reduces
the computing needs to a few CPU years, the light curve 
model-magnitudes and errors must be continuous functions
of redshift, as well as their derivatives.

 
 \section{Results for \SDSS\ and SNLS}
 \label{sec:results_data}

Here we present results for the three-season \SDSS\ data 
and the first-season SNLS data described in \S\ref{sec:data},
and we compare with results from simulations of those same samples.
There are two fit minimizations (\S\ref{sec:zLCFIT_method})
to determine the appropriate filters to include.
The \photoz\ and distance-modulus results are 
taken to be the mean of their respective
probability distribution functions (pdf) marginalized
over the other fit parameters 
using a grid of $11^5$ integration cells
(Appendix~\ref{app:marg}).
The uncertainty is taken to be the rms of the pdf.
A prior on the host-galaxy {\photoz} or SN color
could potentially improve the precision of the method 
and reduce the frequency of 
catastrophic SN \photoz\ outliers;
we have not used such priors here in order to better illustrate
the performance of the \zLCFIT\ method on its own.

\begin{figure}[hb]
\centering
 \epsscale{1.1}
 \plotone{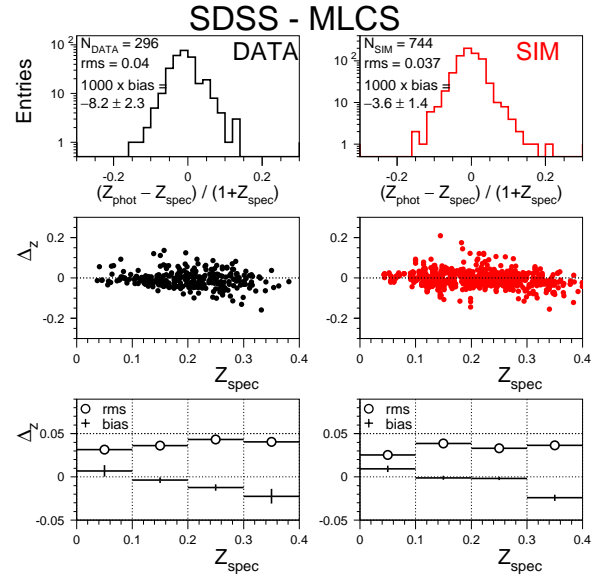}
  \caption{
	Redshift precision for \SDSS\ sample using \zmlcs.
	Left plots are for data; right plots for simulation.
	Top plots show the $\DZSYM$ distribution for the entire sample;
	the number of events, rms, and  bias are indicated on the plot.
	Middle plots show $\DZSYM$ vs. $\Zspec$.
	Bottom plots show the bias and $\zrms$ in redshift bins.
      }
  \label{fig:zresids_SDSS_MLCS}
\end{figure}

\begin{figure}[hb]
\centering
 \epsscale{1.1}
 \plotone{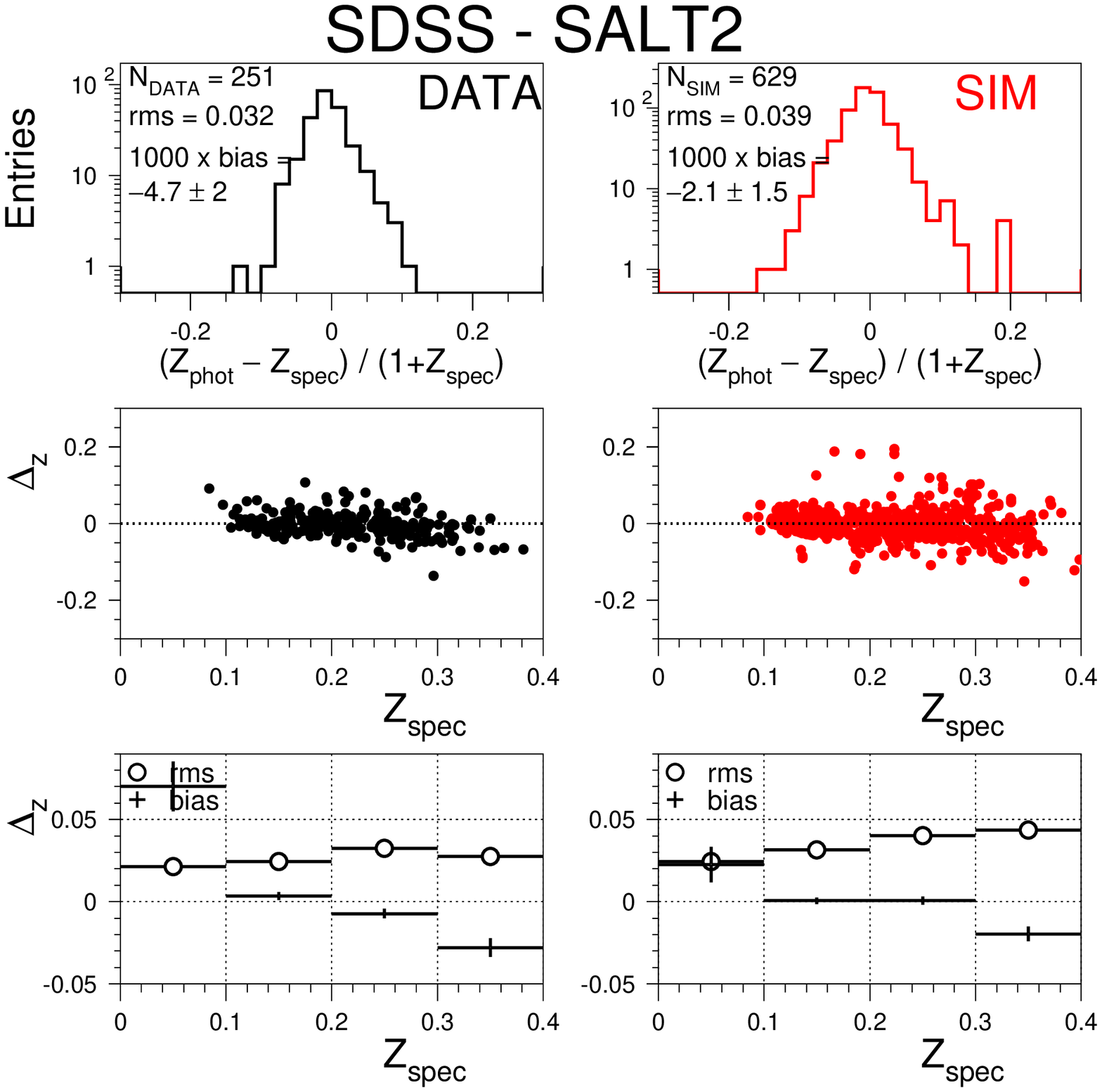}
  \caption{
	Same as Fig.~\ref{fig:zresids_SDSS_MLCS}, 
	but using \zSALTII.
      }
  \label{fig:zresids_SDSS_SALT2}
\end{figure}

\begin{figure}[hb]
\centering
 \epsscale{1.1}
 \plotone{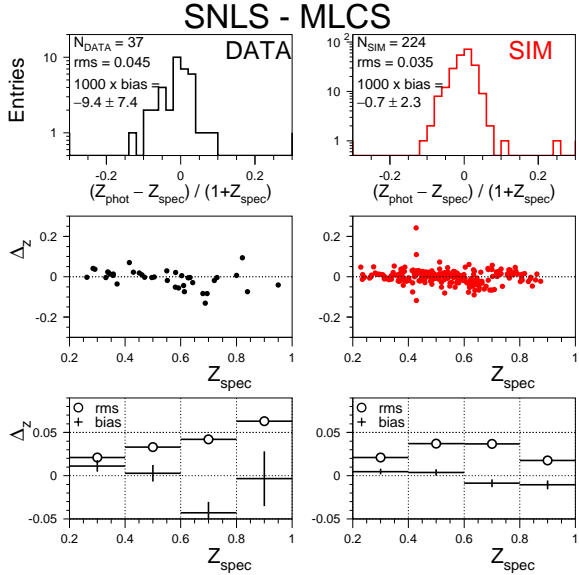}
  \caption{
	Same as Fig.~\ref{fig:zresids_SDSS_MLCS}, 
	but using the SNLS sample.
      }
  \label{fig:zresids_SNLS_MLCS}
\end{figure}

\begin{figure}[hb]
\centering
 \epsscale{1.1}
 \plotone{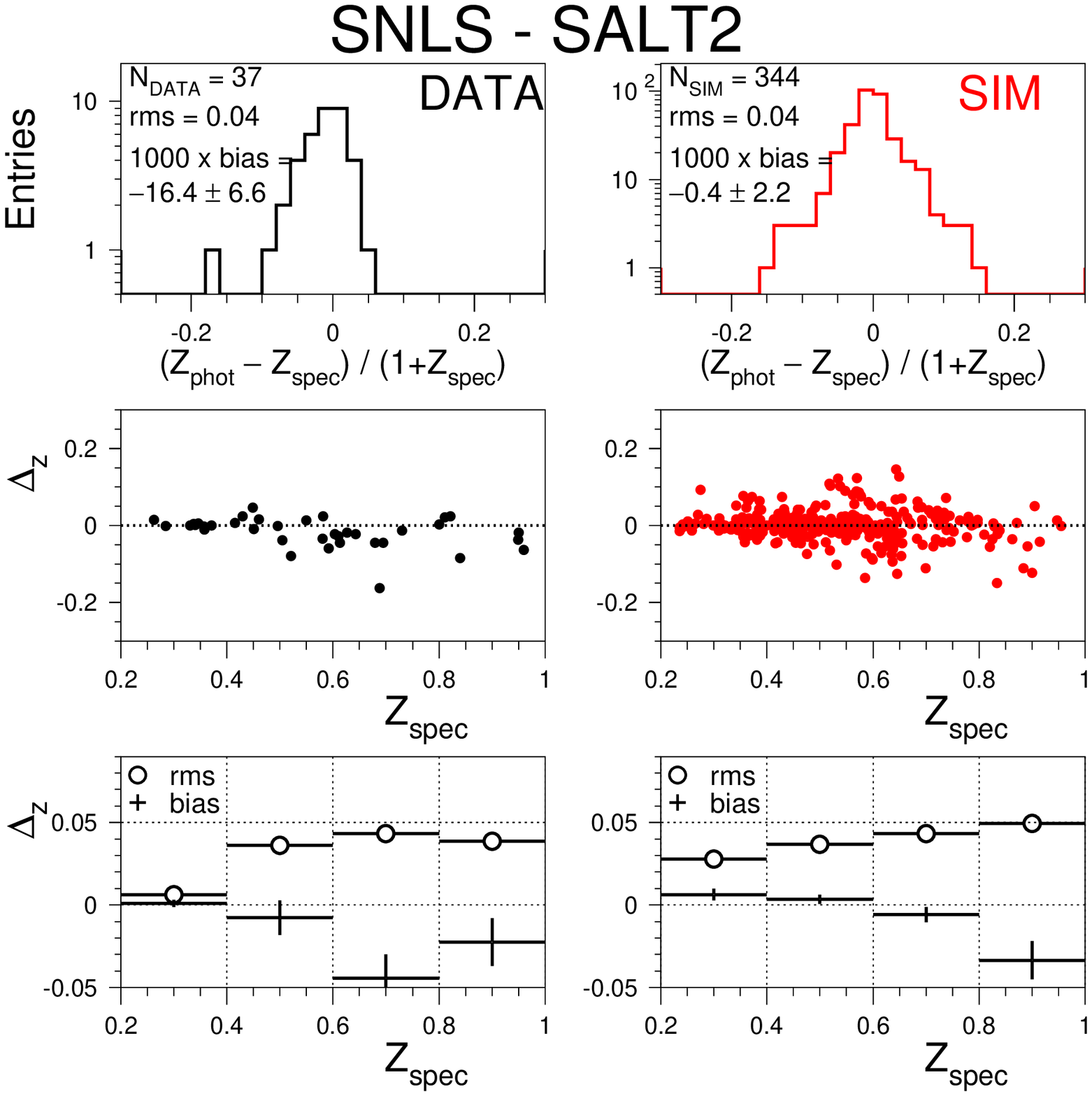}
  \caption{
	Same as Fig.~\ref{fig:zresids_SDSS_MLCS}, 
	but using the SNLS sample and \zSALTII.
      }
  \label{fig:zresids_SNLS_SALT2}
\end{figure}

Following a commonly used practice in the literature, 
we characterize the precision of the \zLCFIT\ \photoz\
precision with the quantity
\begin{equation}
  \DZSYM \equiv \DZDEF ~,
  \label{eq:DZSYMdef}
\end{equation}
and we use $\zrms$ to denote the root mean square of the 
distribution of $\DZSYM$.
The \SDSS\ results are shown in 
Figures~\ref{fig:zresids_SDSS_MLCS} and \ref{fig:zresids_SDSS_SALT2}
for the \zmlcs\ and \zSALTII\ methods, respectively.
There are fewer SNe in the sample fitted with \SALTII\ because
the more restrictive rest-frame model wavelength range (2900-7000~\AA)
rejects $i$-band data for redshifts below about 0.1;
without $i$-band, these low-redshift SNe fail the requirement 
of three observer-frame filters.
The SNLS results are shown in 
Figures~\ref{fig:zresids_SNLS_MLCS} and \ref{fig:zresids_SNLS_SALT2}.
Distributions from the data and simulated samples are shown
side-by-side in these Figures, illustrating the reliability
of the simulations in predicting the dispersions and bias.
Each plot shows the number of SNe satisfying
the selection criteria, $\zrms$ (overall and versus $\Zspec$),
and the redshift bias (overall and versus $\Zspec$).
The $\zrms$ values are $\sim 0.04$ for both the 
\SDSS\ and SNLS samples.

For the \zmlcs\ method applied to the \SDSS\ sample
(Fig.~\ref{fig:zresids_SDSS_MLCS}), the overall $\DZSYM$ 
bias in the data is notably larger than in the simulation,
and there is a redshift-dependent bias that is partially
predicted by the simulation.
There are two contributions to the $\DZSYM$ bias.
First, there is a degeneracy between intrinsic reddening 
(due to extinction or color) and redshift.
For redshifts $\Zspec > 0.3$, the best-fit \photoz\ is sometimes
near $\Zphot \sim 0.15$, and the best-fit color corresponds to 
a very red and intrinsically dim SN~Ia. 
This degeneracy is sensitive to the S/N;
the resulting bias is redshift dependent and
is well modeled by the simulation.

To potentially identify fits with a catastrophic \photoz\ error 
resulting from a strong color-redshift degeneracy,
we have looked for a second maximum in the one-dimensional
marginalized pdf.
For \zmlcs, $\sim 9$\% of the fits have a second maximum in 
both the \photoz\ and color pdf, and for \zSALTII\
the corresponding fraction is $\sim 4$\%. For both models,
the \photoz\ precision is the same for the subset with a
second maximum in the pdf, and therefore this approach cannot
be used to identify \photoz\ outliers in the \SDSS\ sample.
We also find no correlation between the \photoz\ \unc\
and catastrophic outliers.
Clearly a reliable host-galaxy \photoz\ prior will help reduce
catastrophic outliers, and this information will be used in
future analyses that include \specy\ unconfirmed SNe~Ia 
in the Hubble diagram.

The second contribution to the $\DZSYM$ bias is 
related to the $U$-band anomaly discussed in K09,
and this source of bias is not modeled in the simulation.
The sub-sample with $z > 0.2$, where the observer frame 
$g$ band corresponds to the rest-frame $UV$ region, 
has a bias larger than the average.
The sub-sample with $z < 0.2$ has a much smaller bias.

\begin{figure}[hb]
\centering
 \epsscale{1.1}
 \plotone{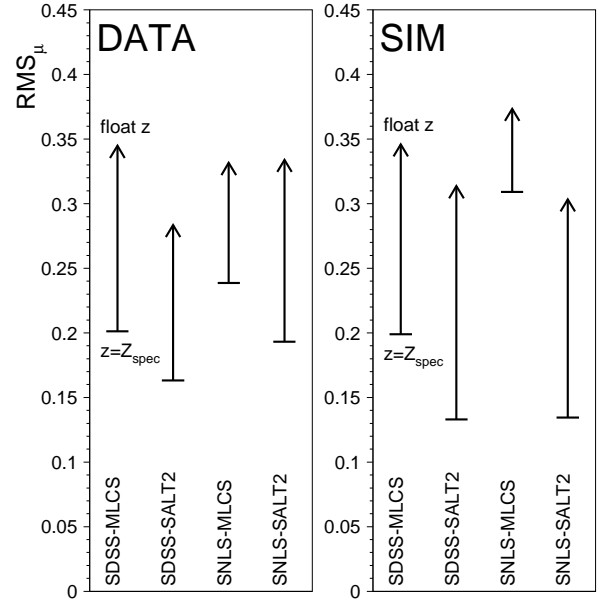}
  \caption{
	$\murms$ for four-parameter \mlcs\ and \SALTII\ fits
	(bottom of each arrow) and for five-parameter 
	\zmlcs\ and \zSALTII\ fits (top of each arrow).
	The left panel shows $\murms$ for the data; 
	the right panel shows the simulations.
      }
  \label{fig:murms}
\end{figure}

For cosmological applications, it is of interest to study how 
the use of {\photoz}'s in place of \spec\ redshifts impacts the 
determination of SN distances.
Distance modulus ($\mu$) dispersions for fits with both 
spectroscopic and photometric redshifts are shown in 
Fig.~\ref{fig:murms}, where $\murms$ is the rms scatter 
of the distribution of $\mufit - \muref$. 
The reference distance modulus ($\muref$) is calculated from 
the \spec\ redshift and the same standard cosmology 
used in the simulation: $w=-1$, $\OM=0.3$, $\OL=0.7$.
For \mlcs, $\mufit$ is the fitted distance modulus. 
For \SALTII, $\mufit$ is defined to be
\begin{equation}
  \mufit^{\rm SALT2} \simeq 30 - 2.5\log_{10}(x_0) + 
         \alpha\cdot x_1 - \beta\cdot c ~,
  \label{eq:muSALTII}
\end{equation}
where $\alpha=0.11$ and $\beta=2.6$ are fixed parameters
from the simulation.
Note that the particular choice of $\alpha$ and $\beta$ does 
not affect the $\mu$ dispersion.
Compared to the four-parameter light curve fits
using \spec\ redshifts, the \zLCFIT\ method increases $\murms$ 
by 0.1 to 0.2~mag.

To gauge the appropriate level of intrinsic magnitude variations 
in the simulation, we compare the $\DZSYM$  precision in the data 
to that in two different simulated samples.
The first sample is generated with 0.1~mag color-smearing,
and the second sample is generated with 0.2~mag coherent smearing.
The width of a Gaussian fit to the 
$\DZSYM$ distribution ($\zsig$) is used instead of $\zrms$ 
to reduce sensitivity to outliers.
The $\zsig$ results are shown in Table~\ref{tb:zsig}
for the \SDSS\ and SNLS data and for the simulated samples.
Using coherent smearing, the simulated \photoz\ precision 
is significantly better than that of the data, 
while the color smearing model matches the data well. 
For the small SNLS data sample fitted with \zmlcs,
the large \unc\ on $\zsig$ is due to a statistical anomaly
in the distribution that results in a poor fit to a Gaussian.

This empirical estimate of random intrinsic color dispersion 
needed in the simulation does not necessarily suggest that 
there are random color variations in SN light curves, 
but rather that there are additional sources of color variation 
that are not captured by the light curve models.
Using the nearby SN~Ia sample ($z<0.1$), \citet{Nobili2007}
also found evidence for intrinsic color dispersion.
They fit the SNe with a light curve model that includes 
many more color parameters than \mlcs\ or \SALTII,
and their estimate of the color dispersion is considerably 
smaller than our empirical estimate based on 
matching the \photoz\ precision.

\begin{table}[hb]
\caption{
  	Photo-$z$ Precision $\zsig$\protect\footnotemark[1] 
           for Data and Simulations
       } 
\begin{center}
\begin{tabular}{l | llll}
\tableline\tableline
         &  \SDSS\    &  \SDSS\     & SNLS      & SNLS \\
 sample  &  (\zmlcs)  &  (\zSALTII) & (\zmlcs)  &  (\zSALTII) \\
\tableline  
DATA               & $\zsigDATASDSSMLCS \pm \zsigeDATASDSSMLCS$
                   & $\zsigDATASDSSSALT \pm \zsigeDATASDSSSALT$
                   & $\zsigDATASNLSMLCS \pm \zsigeDATASNLSMLCS$
                   & $\zsigDATASNLSSALT \pm \zsigeDATASNLSSALT$ \\
SIM\footnotemark[2]
                   & $\zsigSIMSDSSMLCS \pm \zsigeSIMSDSSMLCS$
                   & $\zsigSIMSDSSSALT \pm \zsigeSIMSDSSSALT$
                   & $\zsigSIMSNLSMLCS \pm \zsigeSIMSNLSMLCS$
                   & $\zsigSIMSNLSSALT \pm \zsigeSIMSNLSSALT$ \\
SIM\footnotemark[3]
                   & $\zsigSIMCOHSDSSMLCS \pm \zsigeSIMCOHSDSSMLCS$
                   & $\zsigSIMCOHSDSSSALT \pm \zsigeSIMCOHSDSSSALT$
                   & $\zsigSIMCOHSNLSMLCS \pm \zsigeSIMCOHSNLSMLCS$
                   & $\zsigSIMCOHSNLSSALT \pm \zsigeSIMCOHSNLSSALT$ \\
\tableline  
\end{tabular}
\end{center}
\footnotetext[1]{$\DZSYM \equiv \DZDEF$}
\footnotetext[2]{Uses color-smearing model with $\PEAKMAGSIGMA=0.1$~mag.}
\footnotetext[3]{Uses 0.2~mag coherent mag-smearing.}
\label{tb:zsig}
\end{table}

  \subsection{Comparison with Recent SNLS Photo-$z$ Results}
  \label{subsec:SNLS_compare}

Using data and simulations for the SNLS, we compare our 
\photoz\ precision with recent results from PD09.
They use the \zSALTII\ method in a manner very similar to ours.
The differences between our method and theirs are: 
(1) they use all four $griz$ filters, while we use only 
those filters that correspond to the valid rest-frame 
wavelength range;
(2) their initial parameter scan is in redshift only ($\Delta z=0.1$ bins),
while our initial scan is over a two-dimensional grid of
redshift ($\Delta z=\ZBINSCAN$ bins) and color ($\Delta c =\CBINSCAN$ bins); 
(3) they impose priors on the color and redshift, while we do
not use priors; and
(4) they impose less restrictive light curve selection requirements,
including the addition of photometrically identified SNe~Ia
(i.e., \specy\ unconfirmed SNe).
The trade-off between the use of priors versus selection criteria 
mainly affects the rate of catastrophic \photoz\ outliers.
Our choice of using flat
priors is intended to better illustrate
the performance of the \zSALTII\ method on higher-quality 
light curves. While the relaxed cuts in PD09 have the advantage
of increasing the sample size, using a prior on color 
(or on host-galaxy extinction) requires a detailed understanding 
of the underlying color distribution and survey selection function. 
The optimal choice between priors and selection criteria is not addressed
here and will need further study.

The PD09 sample is based on nearly 300 SNe~Ia corresponding
to the first three seasons of SNLS, while we use the publicly 
available sample from A06. To make the selection criteria 
more similar for this comparison, 
we have relaxed our requirement on the 
maximum S/N: three filters must have at least
one measurement each with S/N$>5$ (instead of 8).
Our modified selection results in 
$\NTOTASTIER$ SNe~Ia:
$\NLOZASTIER$  with $z < 0.45$ and
$\NHIZASTIER$  with $z > 0.45$.

To evaluate the \photoz\ precision, we use the PD09 metric 
$\zresSNLS \equiv 1.48 \times {\rm median}\vert\DZSYM\vert$,
where $\DZSYM$ is defined in Eq.~\ref{eq:DZSYMdef}.
This quantity is much closer to the Gaussian sigma of
$\DZSYM$ than to $\zrms$. To quantify the rate of catastrophic
\photoz\ outliers, we define $\eta_x$ as the fraction of 
SNe with a \photoz\ that satisfies $\vert\DZSYM\vert > x$, 
and we follow PD09 in using $x=0.15$.

Table~\ref{tb:zsig_compare} compares our precision metrics with 
those in PD09. We also give the $\zresSNLS$ breakdown for the 
low-redshift ($\Zspec < 0.45$) and high-redshift ($\Zspec > 0.45$) 
ranges. We caution that these comparisons are based on different 
data samples, different selection criteria, and different priors.
For the data comparison, the two analyses are reasonably
consistent in both $\zresSNLS$ and $\eta_{0.15}$ 
for both redshift ranges.
For the simulation comparison there is a subtle disagreement.
The PD09 simulation, which uses coherent mag-smearing,
underestimates the scatter in the low-redshift range  
but accurately predicts the precision
in the high-redshift range.
Our simulation using coherent mag-smearing underestimates 
$\zresSNLS$ (in PD09) for both redshift ranges, 
but our simulation based on color-smearing predicts 
$\zresSNLS$ fairly well, perhaps with a slight overestimate 
of the scatter. 
Our simulation supports
our earlier conclusion that color-smearing is needed to model
the \photoz\ precision; the PD09 simulation supports
our conclusion in the low-redshift range but not in the 
high-redshift range.
Finally, our simulation underestimates the fraction of catastrophic
outliers ($\eta_{0.15} \sim 0.002$), while the PD09 simulation gives good 
agreement with the data ($\eta_{0.15} \sim 0.01$).

\begin{table}[hb]
\caption{
  	\photoz\ Precision for SNLS Data and Simulation.
	Results for PD09 and this work are shown.
	Our results include the \unc\ in parentheses.
       } 
\begin{center}
\begin{tabular}{l | ccc }
\tableline\tableline
               & $\zresSNLS$     & $\zresSNLS$     & $\eta_{0.15}$\footnotemark[1] \\
 Reference     & ($\Z < 0.45$) & ($\Z > 0.45$) & (all $\Z$)  \\
\tableline 
PD09 using     &           &          &         \\
~~DATA\footnotemark[2] (3 seasons)
               &  0.016    &  0.025   &  0.014    \\
~~SIM\footnotemark[2]
               &  0.006    &  0.027   &  0.010    \\
\tableline  
this work using     &           &          &         \\
~~A06 DATA       & 0.005(5) & 0.036(7) &  0.02(2) \\
~~SIM(coherent smear)    & 0.004(2) & 0.016(1) & 0.002(2) \\
~~SIM(color-smear)       & 0.019(3) & 0.030(3) & 0(2)  \\
\tableline  
\end{tabular}
\end{center}
\footnotetext[1]{$\eta_{0.15}$ = fraction of SNe with 
                    $\vert \Zphot-\Zspec \vert/(1+\Zspec) > 0.15$.}
\footnotetext[2]{See ${z_{\rm pho}}^b$ columns in Table 1 of PD09. }
  \label{tb:zsig_compare}
\end{table}

\subsection{Photo-$z$ Correlations}
\label{subsec:fitcorr}

\begin{figure}[hb]
\centering
 \epsscale{1.1}
 \plotone{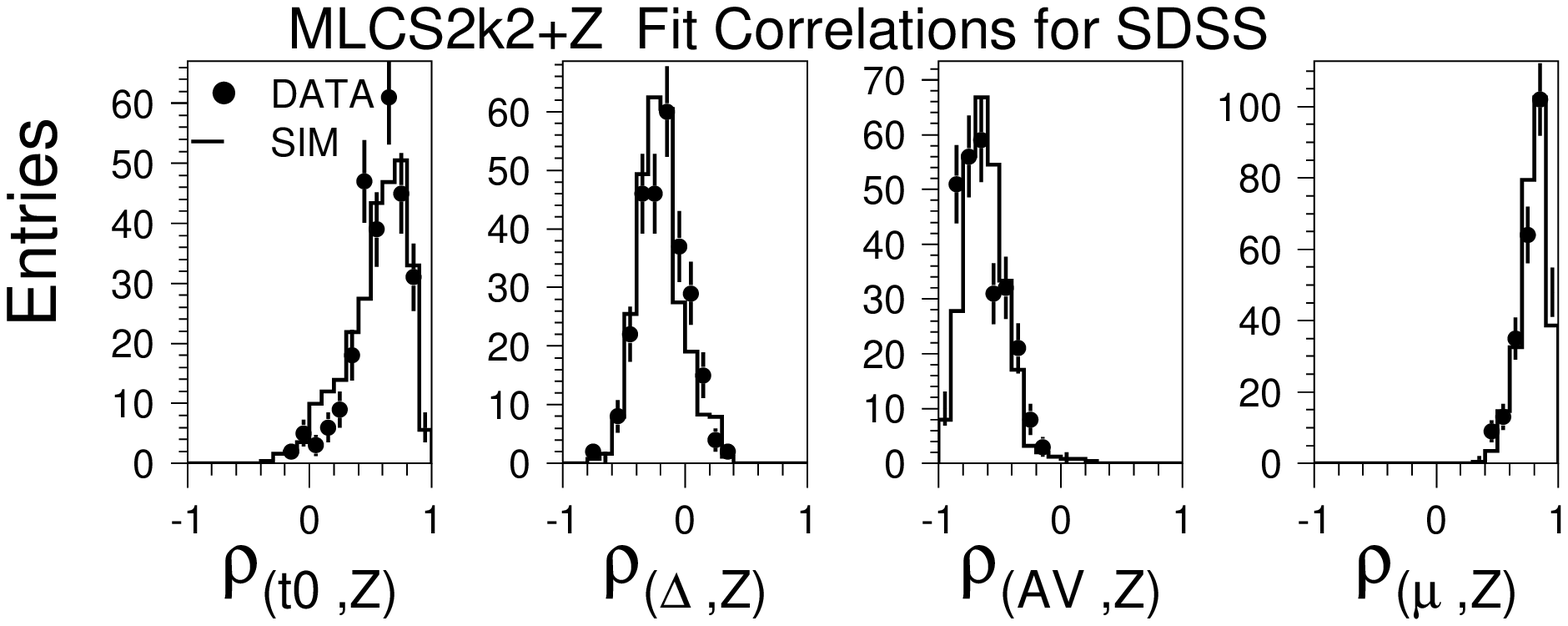}
 \plotone{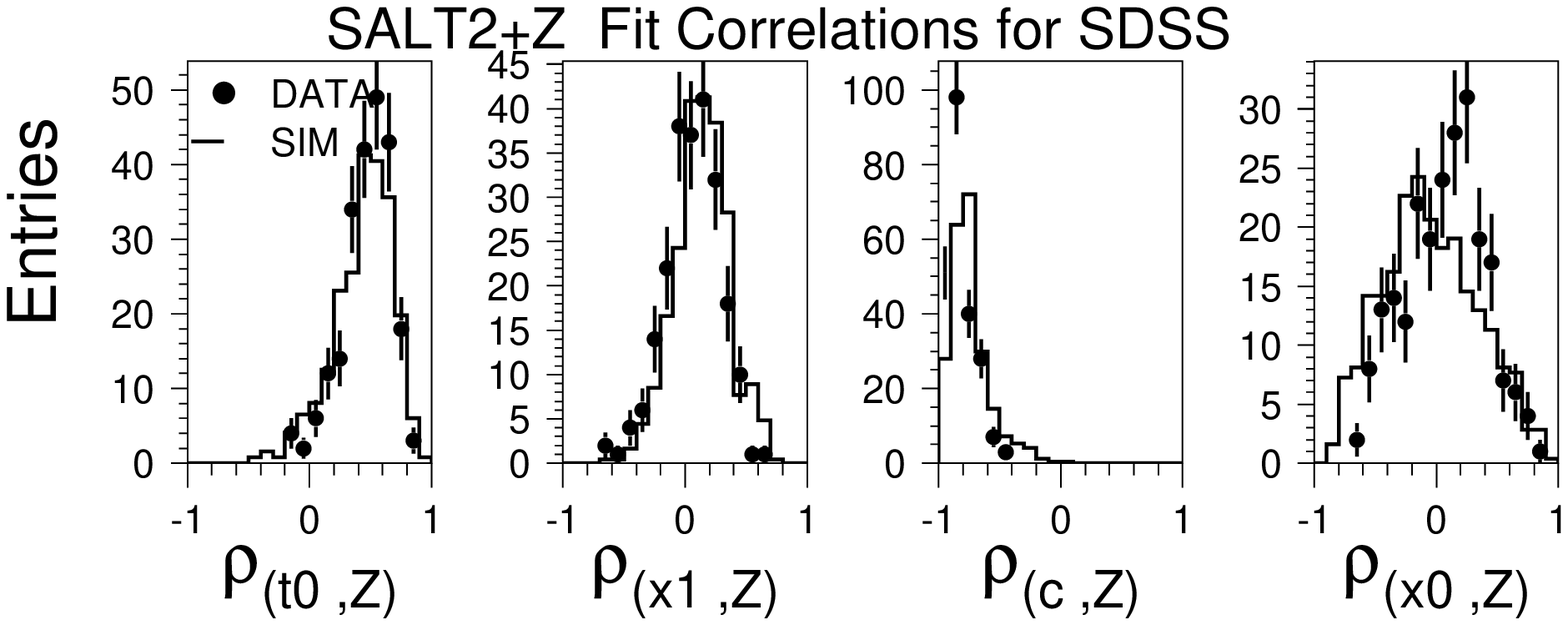}
  \caption{
	Reduced correlations ($\rho$) between the \photoz\ ($Z$)
	and the other four light curve fit parameters as indicated 
	in the title of each plot.
	Top plots are for the \SDSS\ sample using
	\zmlcs; bottom plots are for the \zSALTII\ method.	
	The data are shown by dots; the simulation is shown
	by the histogram.
      }
  \label{fig:fitcorr_SDSS}
\end{figure}

Here we briefly discuss \photoz\ correlations that
should be propagated in a Hubble diagram analysis.
Using the \zLCFIT\ results from the \SDSS\ sample,
Fig.~\ref{fig:fitcorr_SDSS} shows reduced correlations
($\rho$) between the \photoz\ and each of the 
other four light curve fit parameters.
The \zmlcs\ \photoz\ correlation with $t_0$ and distance modulus 
(upper plots) are both peaked at large positive values,
the correlation with extinction ($A_V$) is negative,
and there is little correlation with the shape parameter $\Delta$.
For \zSALTII\ (lower plots) the \photoz\ correlations with $t_0$,
shape/stretch parameter ($x_1$), and color are qualitatively similar
to those based on the \zmlcs\ method.
The $\rho_{(x0,\Z)}$ correlation has a very broad distribution
with an average near zero.
For \zmlcs, the simulated distributions match the data well,
while for \zSALTII\ there is a slight discrepancy in the
distributions of $\rho_{(c,\Z)}$ and $\rho_{x0,\Z}$.
This discrepancy occurs only for $\Zspec > 0.2$ and could 
be an artifact of the subset with smaller S/N.

\section{Predictions for LSST}
\label{sec:lsst}

\newcommand{\OPSIMVERSION}{OPSIM1.29}
\newcommand{\ZPTSYM}{{\rm Z_{\rm p.e.} }}
\newcommand{\mfive}{{\cal M}_{5\sigma ps}}
\newcommand{\msky} {{\cal M}_{\rm sky}}
\newcommand{\ffive}{{\cal F}_{5\sigma ps}}
\newcommand{\fsky}{{\cal F}_{\rm sky}}
\newcommand{\feff}{\epsilon_A}
\newcommand{\SNR}{{\rm SNR}}

Since we have demonstrated that the \snana\ simulation can 
be used to reliably determine the \photoz\ precision for 
the \SDSS\ and SNLS samples, we now turn our attention to
forecasts for the  LSST survey.
LSST will discover far more SNe than \spec\ resources can target, 
and photometric methods will be needed to determine both 
the redshift and SN type for the majority of events.  
Here we determine the precision (bias and rms) on the
\photoz\ and distance modulus using 
the \zLCFIT\ method on more than $10^4$ simulated SNe~Ia 
(after selection requirements) corresponding to the 
LSST-DEEP and LSST-MAIN surveys. 
Contamination from non-Ia SNe and the resulting precision in 
cosmological parameters will be presented in a future work.

The DEEP survey comprises seven fields, each covering nearly 
$10~{\rm deg}^2$, that is densely time-sampled in 
all of the $ugrizY$ LSST filters. 
After rejecting passbands with invalid $\lamf/(1+\Zphot)$,
the average number of observations per SN~Ia
used in the light curve fit is 66.
The MAIN survey covers more than $20,000~{\rm deg}^2$
but is not optimized for SN observations, so 
the light curve sampling often has large temporal 
gaps for each filter. 
The mean number of fitted observations per SN for the 
MAIN survey is 20, more than a factor of 3 fewer
compared to the DEEP fields.
More details about the LSST are given in \citet{Ivezic_08,LSSTSB09}. 
Due to the extensive computing resources needed to 
marginalize these large LSST samples, we have only performed
the minimizations that are adequate for determining central
values for the fitted parameters.

To simulate observing conditions, we use the output of version
{\OPSIMVERSION} of the
LSST Operations Cadence Simulator 
(\citet{OPSIM2006} and \S 3.1 of \citet{LSSTSB09}).
for the cadence, sky noise, and $5\sigma$ limiting magnitude
for each measurement
For each observation, the \snana\ simulation requires a 
zero point ($\ZPTSYM$) to translate the simulated SN magnitude ($m$)
into an observed CCD flux measured in photoelectrons, 
$F = 10^{-0.4(m-\ZPTSYM)}$. In terms of the 
\OPSIMVERSION\ parameters, we calculate this zero point to be
\begin{eqnarray}
  \ZPTSYM & = & 
    2\mfive - \msky + 2.5\log_{10}(A\cdot (S/N)^2)
    \nonumber \\
   & + & 
     2.5\log_{10}\left[
       1 + A^{-1}\times 10^{0.4(\msky-\mfive)} 
               \right] ~,
   \label{eq:ZPT}
\end{eqnarray}
where $\mfive$ is the $5\sigma$ limiting magnitude,
$\msky$ is the Perry sky brightness (mag/arcsec$^2$),
$A= [2\pi\int[{\rm PSF}(r)]^2rdr]^{-1} = (1.51\cdot {\rm FWHM})^2$
is the effective aperture (in arcsec$^2$) where 
FWHM describes the seeing, and S/N$=5$ is the 
signal-to-noise ratio corresponding to $\mfive$.

\begin{figure}[hb]
\centering
 \epsscale{1.05}
 \plotone{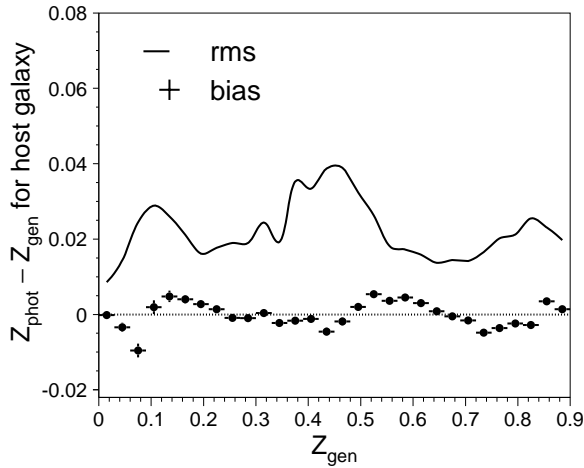}
  \caption{
	For $6\times 10^4$ simulated LSST host galaxies, 
	BPZ bias (crosses) and rms (curve) of $\Zphot-\Zgen$
	vs. the true host-galaxy redshift ($\Zgen$) 
	in redshift bins of width $\Delta z=0.03$.
      }
  \label{fig:LSSTHOST_photoz_resids}
\end{figure}

For this study, we carry out the 
light curve fits both with and without a 
host-galaxy \photoz\ prior;  no other priors are used.
The host-galaxy prior is determined from the 
Bayesian Photometric Redshift Estimation (BPZ) technique
\citep{Benitez2000}
applied to a preliminary set of simulated galaxies.
The galaxy colors and luminosities are generated to match
observed distributions as a function of redshift.
For more details, see \S 3.8 of \citet{LSSTSB09}.
The signal to noise as a function of apparent magnitude
for the host-galaxy simulation is based on co-added exposures 
for 10 years of the MAIN survey, and \photoz's are 
determined for galaxies with $r$ magnitudes down to 25.
The average host-galaxy \photoz\ precision from BPZ is 0.02, 
and there are some variations with redshift as shown in 
Fig.~\ref{fig:LSSTHOST_photoz_resids}.
The $\Zphot-\Zgen$ bias versus redshift has wiggles of order 0.005,
but the true bias could be larger if the SN host galaxies are 
not a random subset of the galaxies used for \photoz\ training.
Note that in this LSST discussion of the host galaxy and SNe~Ia,
we characterize the \photoz\ precision in terms
of $\Zphot-\Zgen$ instead of $\DZSYM$.

These host-galaxy \photoz\ values are stored in a library
for the \snana\ simulation.
For each simulated SN with true redshift $\ZSN$, 
the host galaxy with true redshift ($\Zgal$)  closest to $\ZSN$
is selected. The corresponding host-galaxy \photoz\ is then scaled
by the ratio  $\ZSN/\Zgal$ to correct for the slight redshift
mismatch between the SN and the host galaxy.
The scaled host-galaxy \photoz\ and its \unc\
are used to impose a Gaussian prior in \zLCFIT.


\begin{figure*} 
\centering
 \epsscale{.5}
 \plottwo{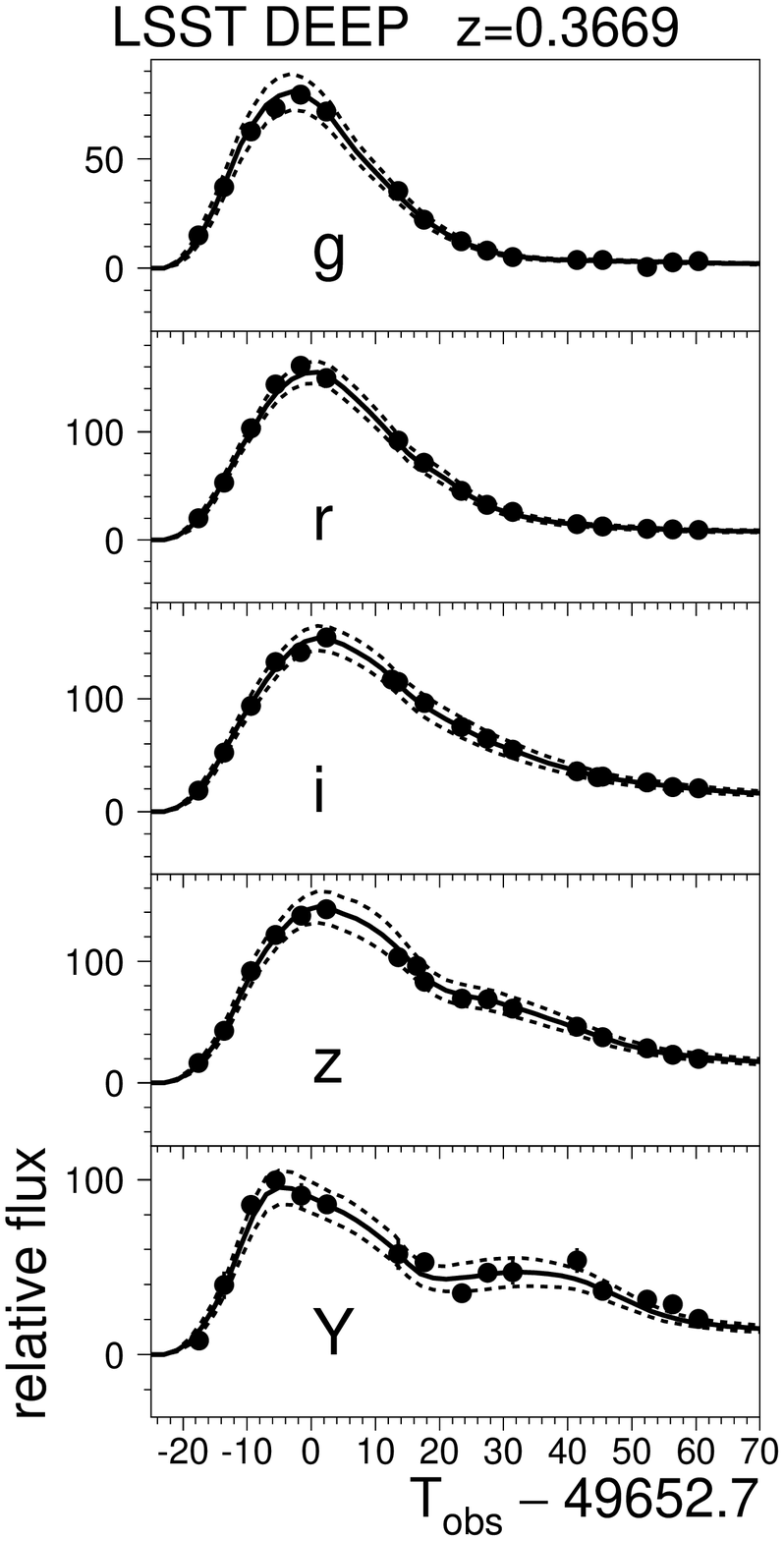}{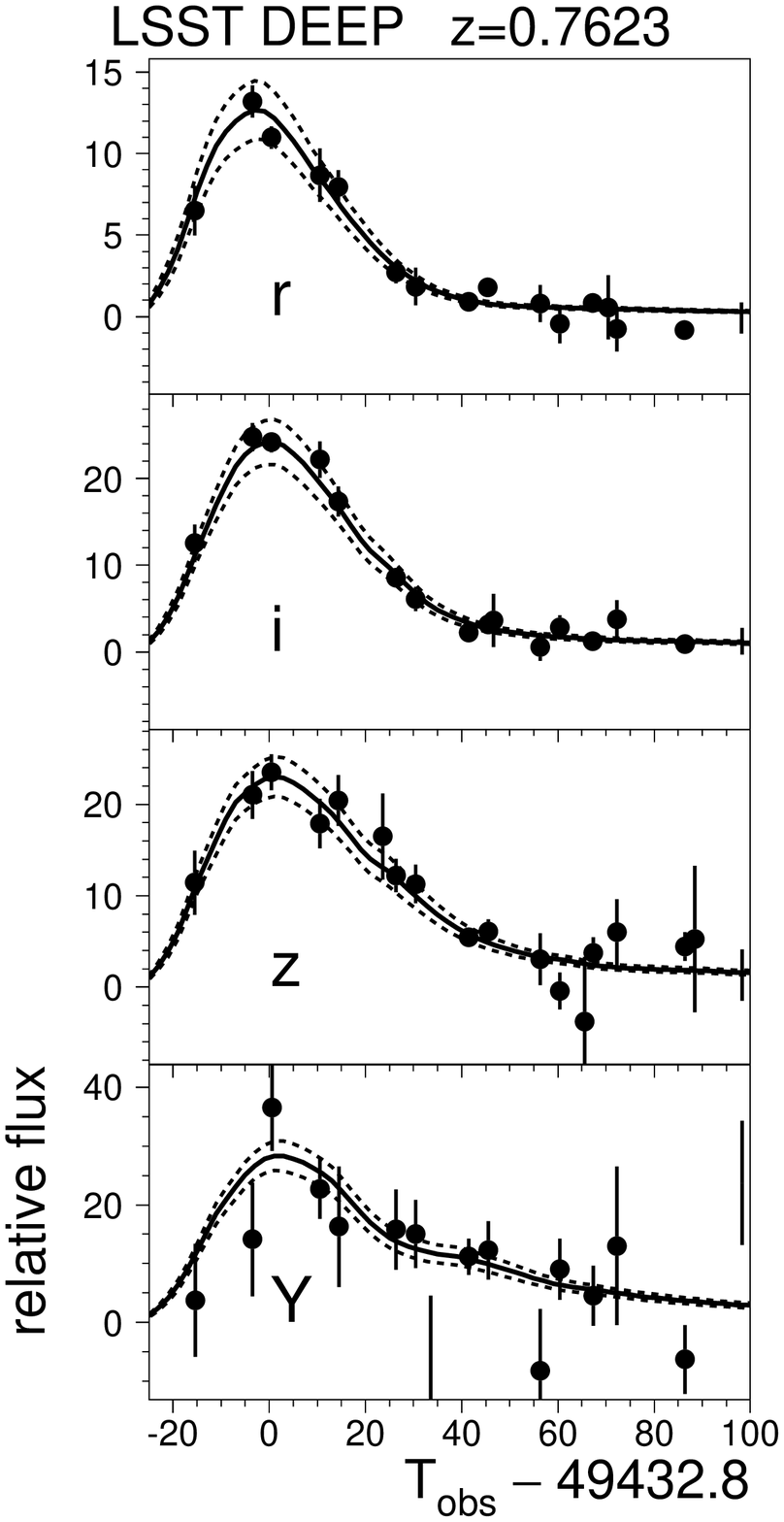}
 \plottwo{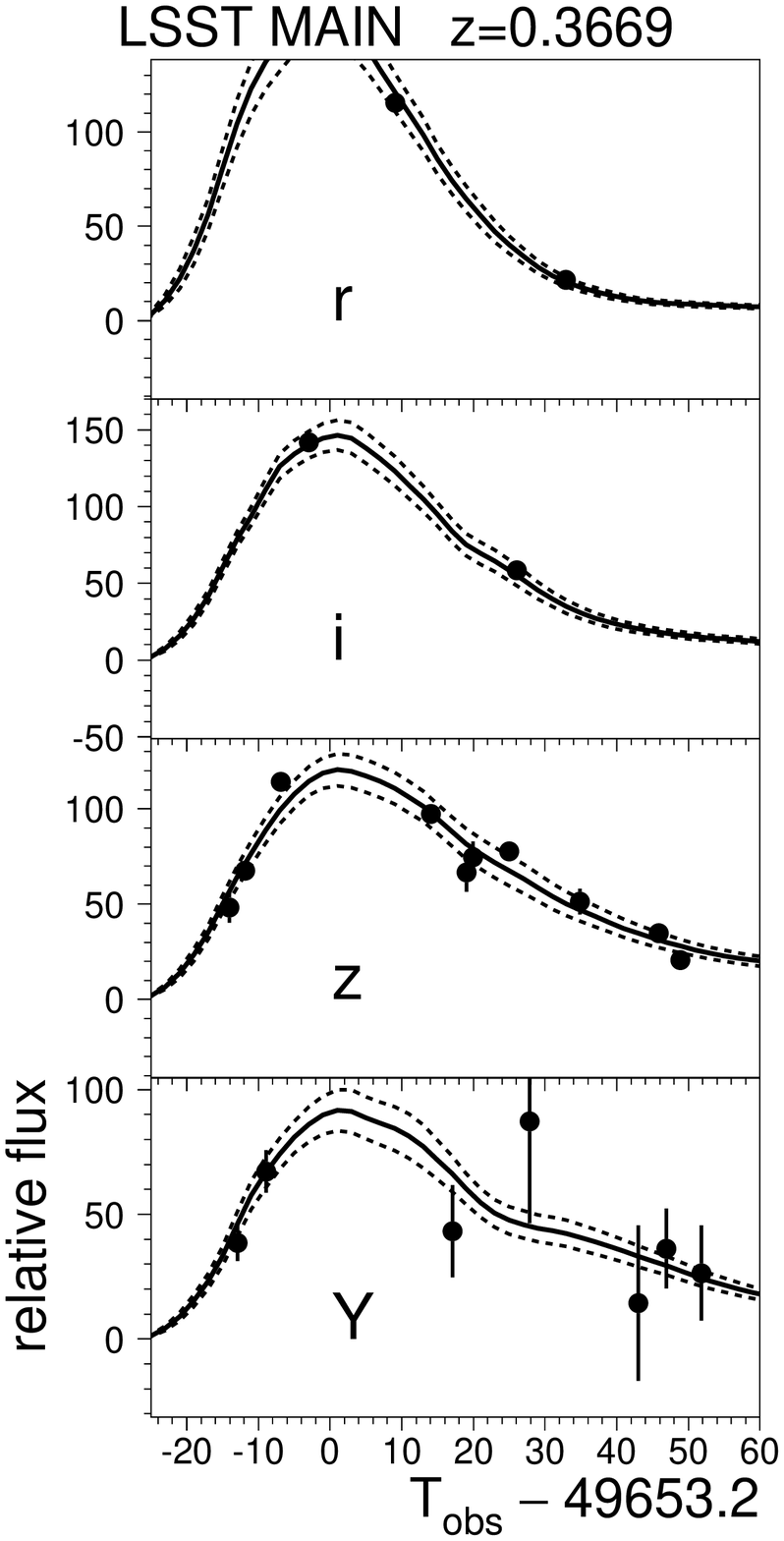}{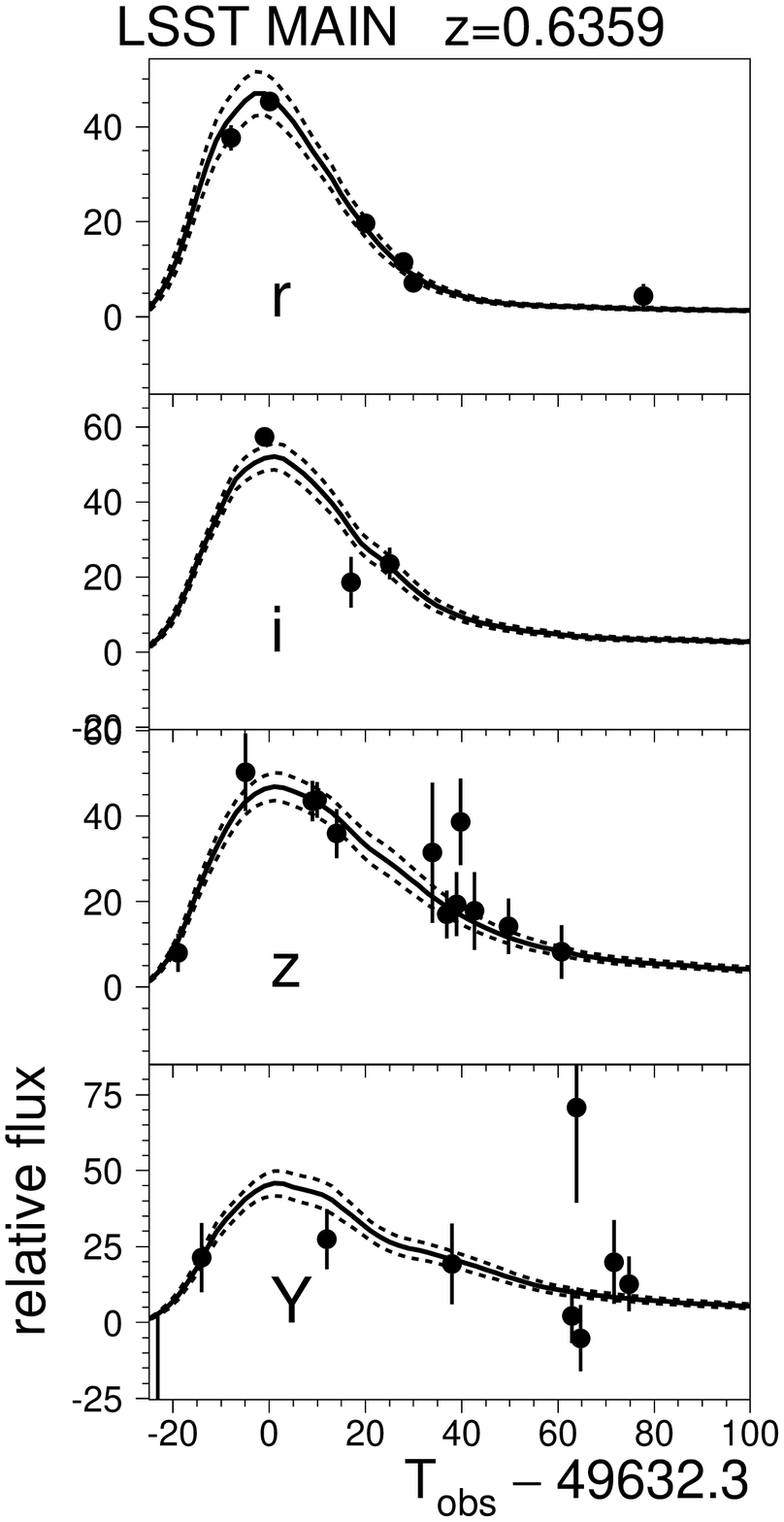}
  \caption{
	Left two panels show typical SN~Ia light curves simulated 
  	for the LSST-DEEP fields;
	right two panels show typical light curves 
	for the LSST-MAIN fields.
	The redshift is indicated on the top of each plot. 
	Dots are simulated fluxes,
	the solid curve is the best-fit \zmlcs\ model,
	and the dashed curves
	are the $\pm 1\sigma$ error bands for the model.
      }
  \label{fig:LSST_lcplot}
\end{figure*}

To ensure well-sampled light curves for the \zLCFIT\ method, we apply 
the following selection requirements to the simulated LSST SN data:
(1) at least two filters with a measurement satisfying $\Trest < -5$ days;
(2) at least two filters with a measurement satisfying $\Trest > +20$ days;
(3) largest rest-frame gap (that overlaps $-5$ to $+20$ days) is $<15$ days;
(4) at least three observer-frame filters have an epoch with 
${\rm S/N}>10$; and 
(5) the light curve fit probability satisfies $\PROBCHI > 0.02$.
The requirement of at least two filters (cuts (2) and (3))
removes poorly sampled light curves predominantly from
the MAIN survey. 
Using these requirements, the  number of SNe per year is 
$\NCUTSLSSTDEEP$ and  $\NCUTSLSSTMAIN$ for the 
DEEP and MAIN surveys, respectively.
We note that these selection requirements are based on educated guesses
rather than an optimization procedure.
Example light curves from the DEEP and MAIN surveys are shown in
Fig.~\ref{fig:LSST_lcplot}.

As a test of the fitting software, we first simulate 
ideal DEEP-field samples with no intrinsic mag-smearing
and the exposure time artificially increased by a factor 
of $10^4$ compared to the nominal exposure.
The resulting \photoz\ bias is less than 0.001 at all redshifts
for both \zmlcs\ and \zSALTII, and the \photoz\ dispersion (rms) 
is less than 0.003. 
The bias on the distance modulus varies between 
0 and 0.01~mag for \zmlcs, and is less than 
$\pm 0.005$~mag for \zSALTII.
The distance modulus dispersion is $\sim 0.02$~mag 
for both fitting models.

  \subsection{Results for LSST Simulations}
  \label{subsec:lsst_results}

\begin{figure*}
\centering
 \epsscale{1.}
 \plottwo{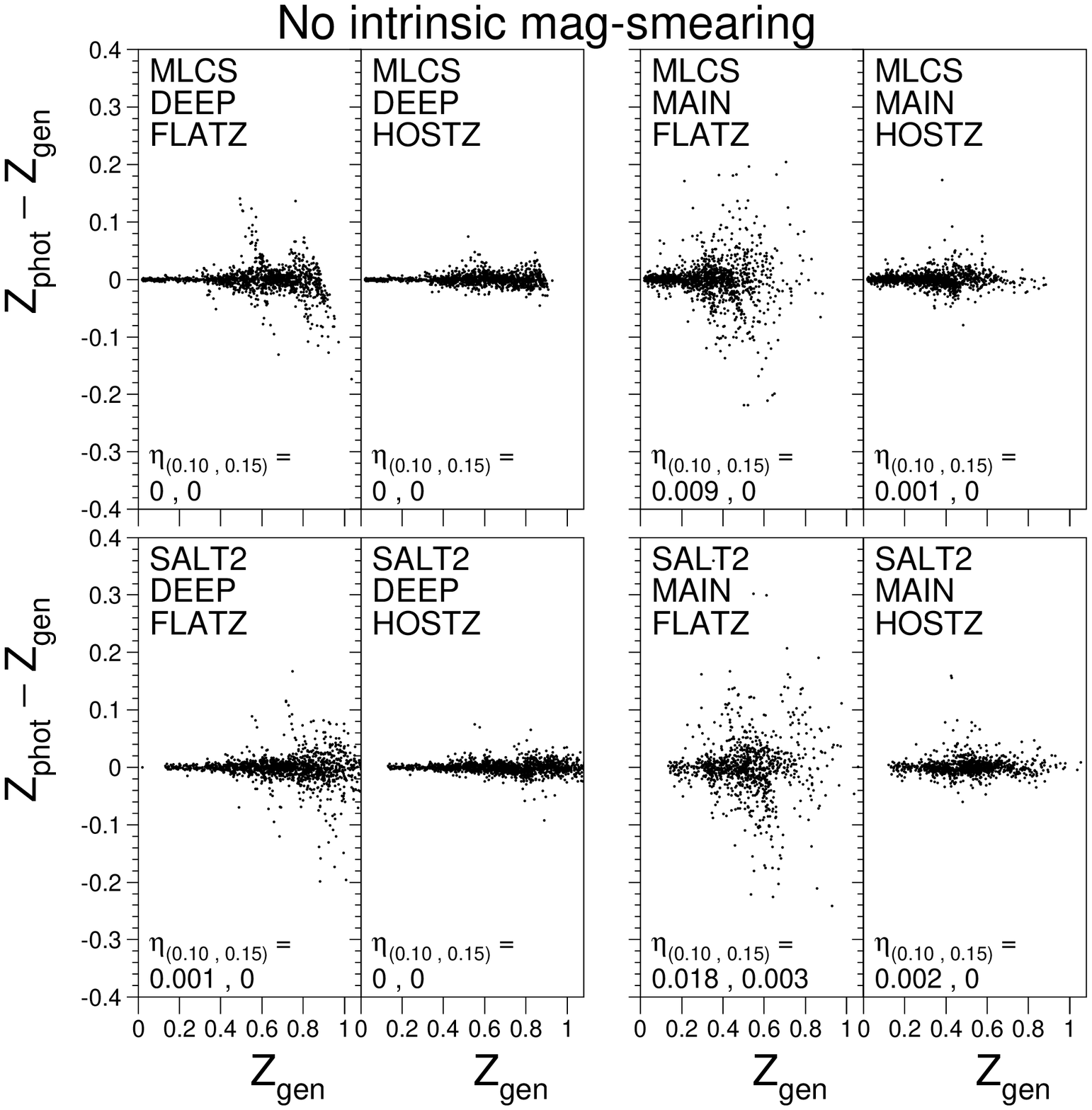}{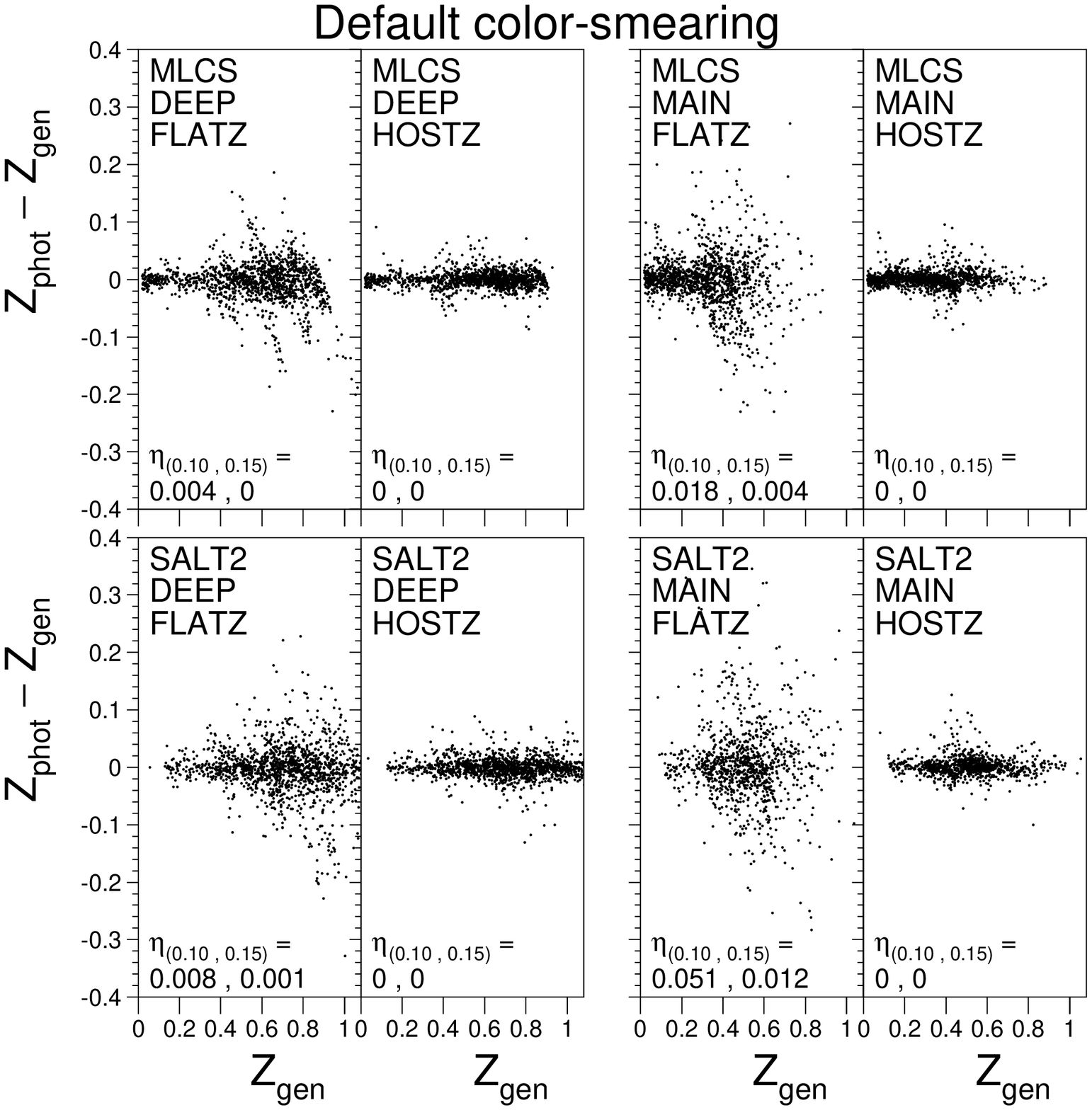}
  \caption{
	$\Zphot - \Zgen$ residuals vs. $\Zgen$ for the
	model (\mlcs or {\SALTII}), survey (DEEP or MAIN), and
        \photoz\ prior (host galaxy or flat) indicated
	in each panel for LSST simulations.
	Each pair of plots compares
	residuals with no \photoz\ prior (FLATZ) to 
	residuals with host-galaxy \photoz\ prior (HOSTZ).
	The fraction of catastrophic outliers ($\eta_{0.10~,~0.15}$),
        indicated in each panel, 
	is defined in the text.
	The simulations are generated without intrinsic
	mag-smearing (left plots) and with the default 
	color-smearing model (right plots).
      }
  \label{fig:LSST_z2d}
\end{figure*}

The SN \photoz\ residuals as a function of redshift
are shown in Fig.~\ref{fig:LSST_z2d} for 
a simulation without intrinsic magnitude fluctuations and 
for a simulation using the same intrinsic color fluctuations 
needed to match the \photoz\ precision
for the \SDSS\ and SNLS data samples
(\S\ref{sec:results_data}).
Each panel shows the residuals as a function
of fitting method (\zmlcs\ or {\zSALTII}), 
survey field (DEEP or MAIN) and 
redshift prior (flat or host-galaxy {\photoz}).
With no intrinsic fluctuations, 
the most notable effects are: 
(1) with a flat redshift prior, the extreme \photoz\ 
outliers extend up to $|\Zphot - \Zgen| \sim 0.2$, 
and (2) the host-galaxy \photoz\ prior significantly reduces 
the number of outliers. 
With default color fluctuations, 
there is a notable increase in the \photoz\ outliers.
In both cases, the redshift range is higher for {\zSALTII}, 
because it extends to a lower rest-frame wavelength (2900~\AA)  
than \mlcs\ (3200~\AA).

We quantify the rate of catastrophic outliers using $\eta_{0.15}$ 
and  $\eta_{0.10}$ (see \S \ref{subsec:SNLS_compare}).
With $\sim 10^4$ SNe per sample, the approximate \unc\ is
$\sigma_{\eta} \simeq \sqrt{\eta}/100$.
Without intrinsic mag-smearing (Fig.~\ref{fig:LSST_z2d}-left),
$\eta_{0.15}=0$ in all cases. In this case,
for the MAIN survey, $\eta_{0.10} \sim 0.01$
without a host-galaxy \photoz\ prior,  and $\eta_{0.10}$ 
is $\times 10$ smaller when the host-galaxy \photoz\ 
prior is used.
Using the default color-smearing model in the simulation
(Fig.~\ref{fig:LSST_z2d}-right),
and fitting without a host-galaxy \photoz\ prior,
$\eta_{0.15} < 0.001$ for the DEEP survey,
and it is somewhat larger for the MAIN survey:
$0.004$ and $0.012$ using the \zmlcs\ and \zSALTII\ methods, respectively.
For $\eta_{0.10}$, the corresponding fractions are $\times 4$ larger.
Using the \zSALTII\ method, the number of outliers is about
$\times 3$ larger compared to \zmlcs; this difference
could be related to the model, but it could also be an artifact
of our implementation. We therefore make no claims that either
method is more precise or has fewer catastrophic outliers.
When the host-galaxy \photoz\ prior is used,
$\eta_{0.10} = \eta_{0.15} = 0$ in all cases.

\begin{figure*} 
\centering
 \epsscale{1.}
 \plottwo{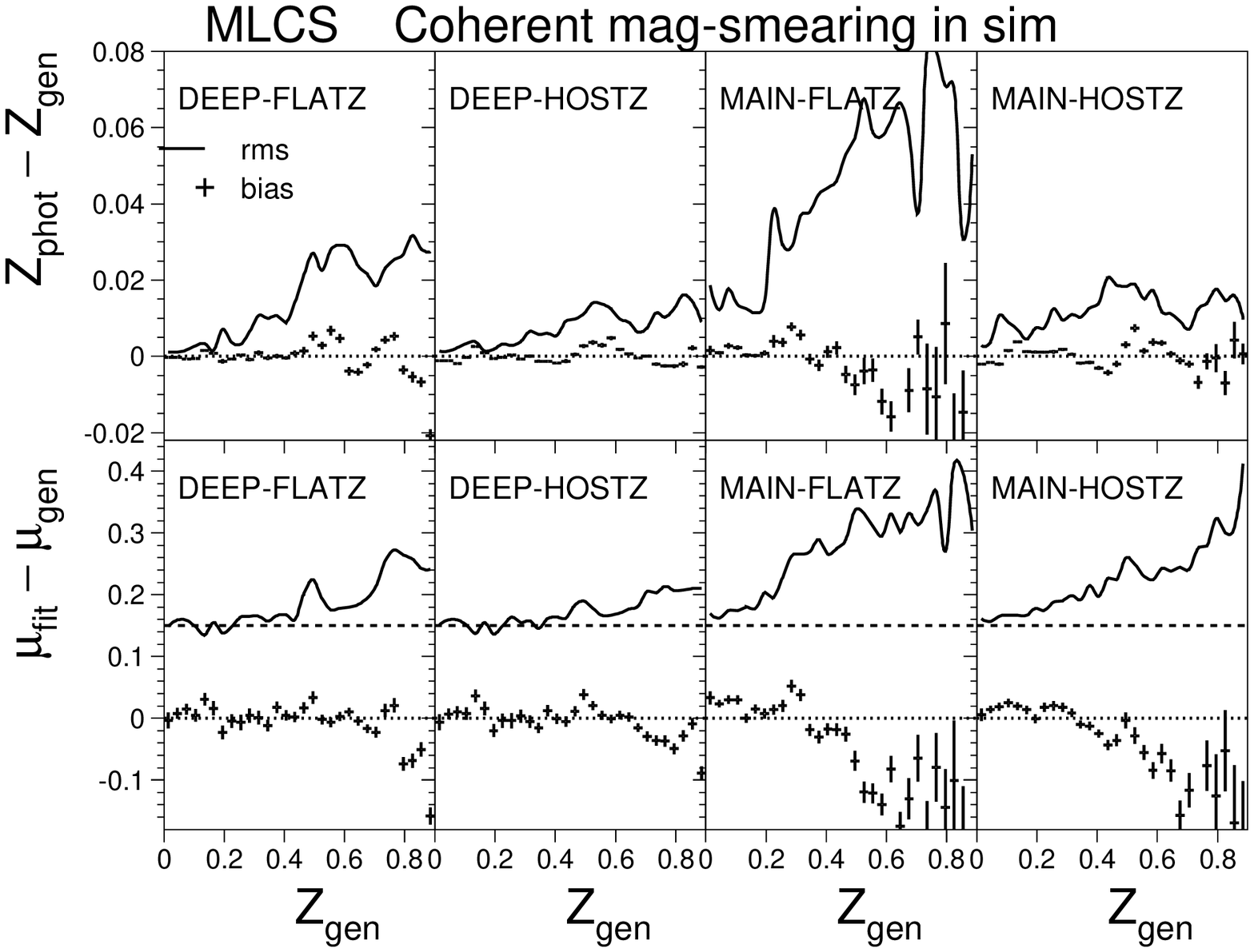}{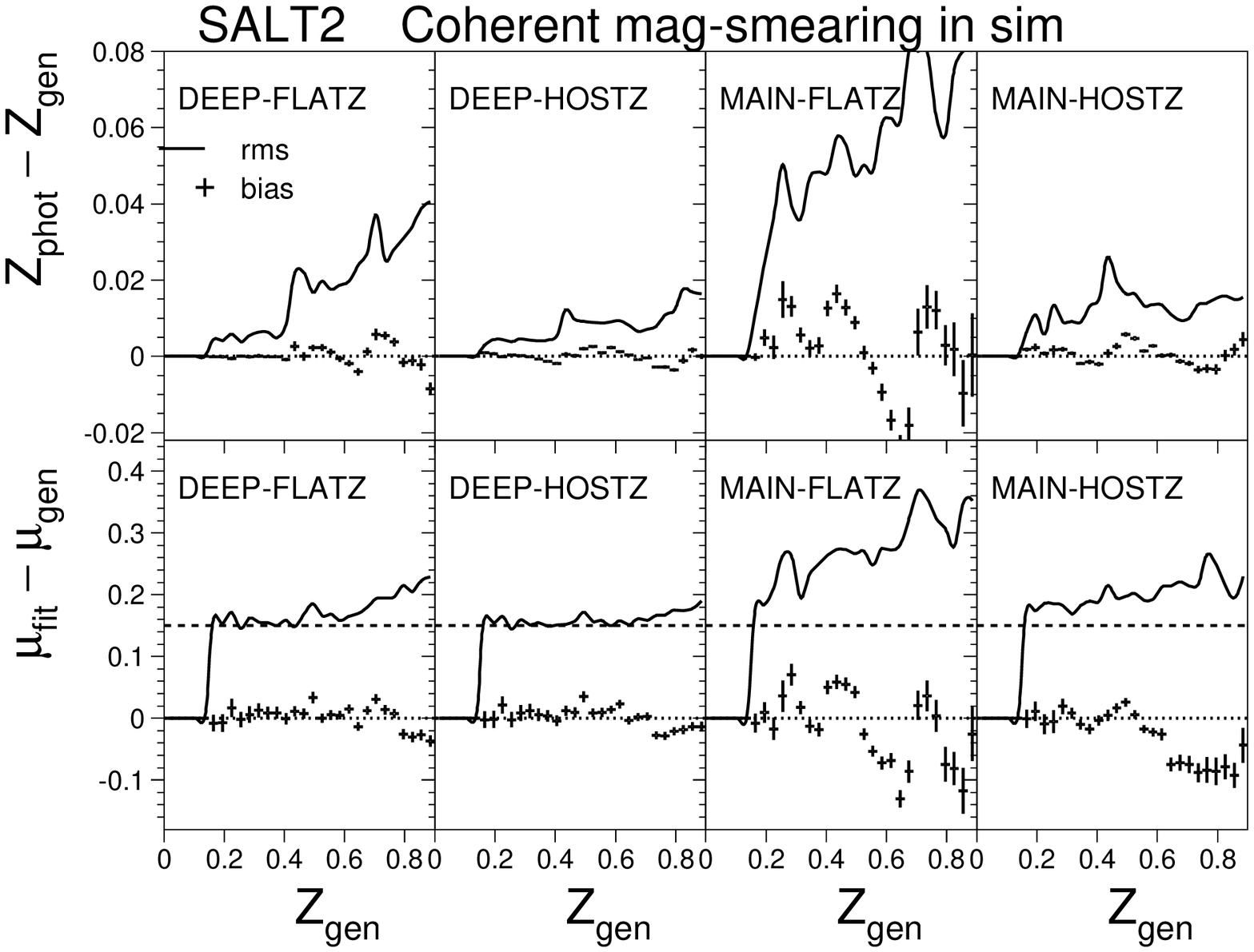}
  \caption{
	Bias (crosses) and rms (curve) on \photoz\ residual 
	($\Zphot-\Zgen$), and on distance modulus residual
	($\mufit-\mugen$) vs. the true redshift ($\Zgen$) 
	in $z$-bins of width $\Delta Z=0.03$ for the LSST simulations.
	Each plot indicates DEEP or MAIN fields, 
	and FLATZ (SN only) or HOSTZ (host-galaxy \photoz\ prior).
	The simulation uses coherent (0.15~mag) brightness 
	fluctuations in all filters and is fitted with 
	\zmlcs\ (left) and \zSALTII\ (right).
	For $z<0.15$, the  \SALTII\ curves drop to zero because
	there are only two valid filters ($gr$), and the data fail
	the three-filter requirement.
	The dashed horizontal lines in the $\mufit-\mugen$ plots
	show the rms contribution from intrinsic variations.
      }
  \label{fig:LSST_COHSMEAR}
\end{figure*}

\begin{figure*}  
\centering
 \epsscale{1.}
 \plottwo{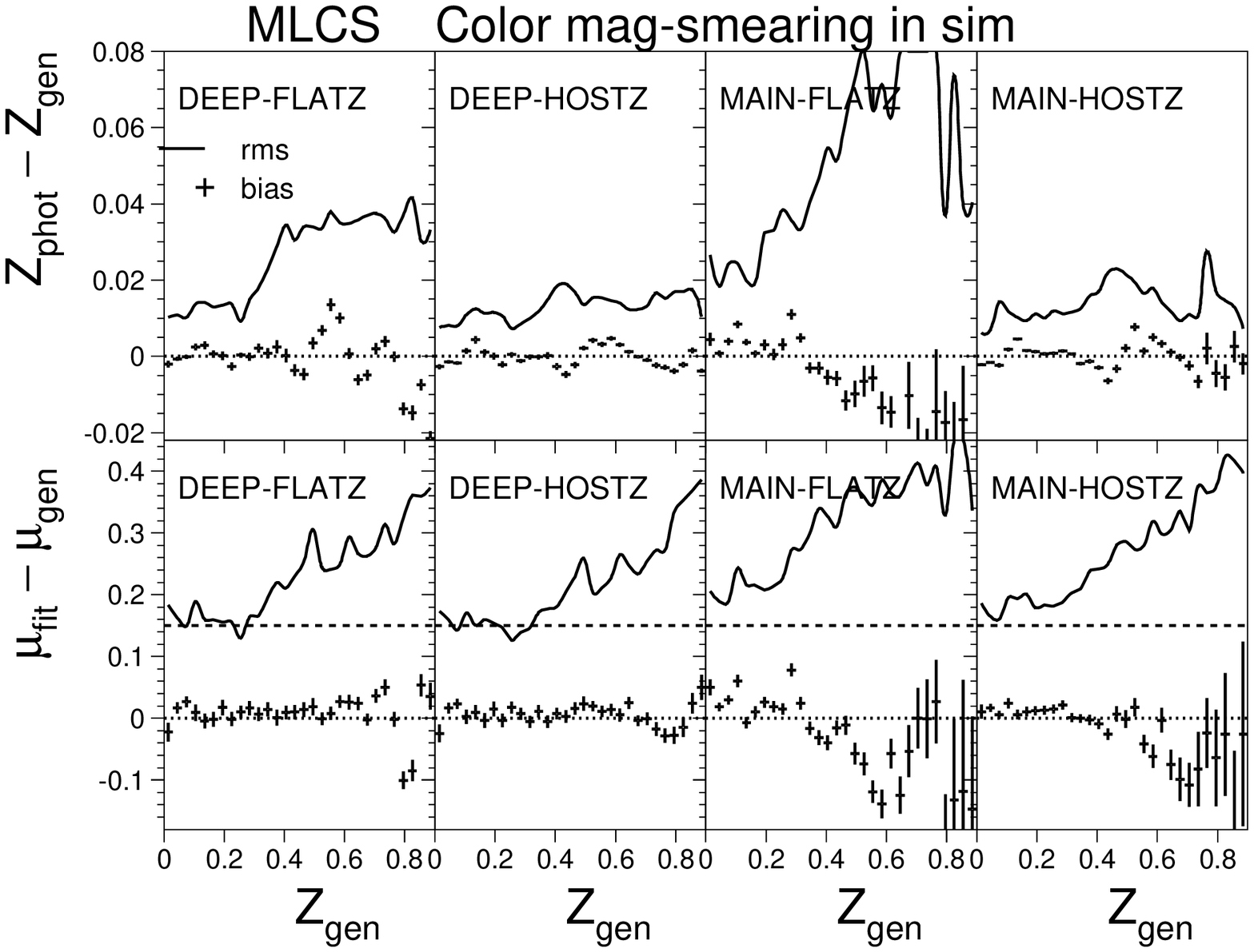}{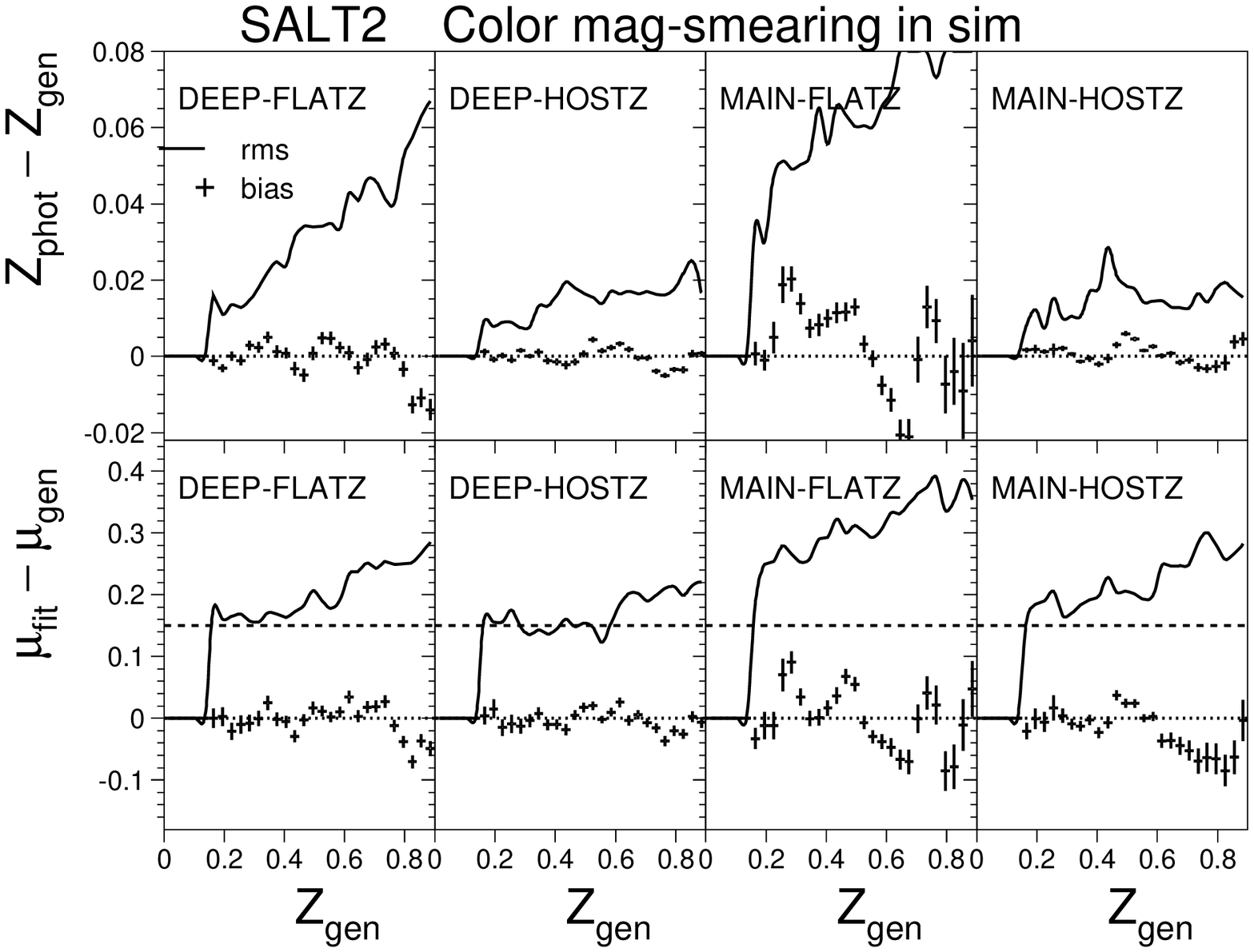}
  \caption{
 	Same as Fig.~\ref{fig:LSST_COHSMEAR}, except that the
	default color-smearing model is used in the simulation.
      }
  \label{fig:LSST_COLORSMEAR}
\end{figure*}

To quantify the precision of the \zLCFIT\ method,
we have evaluated the bias and rms spread as a function
of redshift for both the \photoz\ and distance modulus ($\mu$).
Figure~\ref{fig:LSST_COHSMEAR} shows the results
based on simulations using the coherent mag-smearing model,
and Fig.~\ref{fig:LSST_COLORSMEAR} shows the results
using the default color-smearing model.
The main differences between using these two models
of intrinsic variations are:  
(1) the \photoz\ rms goes to nearly zero at 
low redshift for the coherent mag-smearing model, 
but has a floor of about 0.01 for the color-smearing model,
and (2) the rms is slightly larger at high redshift 
for the color-smearing model.

Here we briefly summarize the precision based on simulations 
using the default color-smearing model and fitting without a 
host-galaxy \photoz\ prior 
(see ``FLATZ'' panels in Fig.~\ref{fig:LSST_COLORSMEAR}).
In the DEEP survey, the \photoz\ rms precision is $\sim 0.01$
at low redshifts and rises to about 0.04 at the highest redshifts. 
The $\murms$ precision is near the 0.15~mag floor at low redshifts 
and roughly doubles at the highest redshifts.
In the MAIN survey, the \photoz\ precision is about
$\times 2$ worse compared to the DEEP field survey.
The corresponding $\murms$ precision is about 0.2~mag at 
the lowest redshifts and also roughly doubles at the highest redshift.
The \photoz\ bias in the DEEP survey has 
wiggles of amplitude $\sim 0.01$ as a function of redshift. 
The $\mu$-bias wiggles are 
at most at the 0.01~mag level, and are notably less
apparent than those seen in the \photoz\ bias.
In the MAIN survey, the bias
is noticeably larger and redshift-dependent;
the \photoz\ bias reaches 0.02 
and the $\mu$-bias reaches 0.1~mag.
Both the \photoz\ and $\mu$ biases are largest
at $\Zgen\sim 0.6$.

When fitting with a host-galaxy \photoz\ prior
(see ``HOSTZ'' panels in Fig.~\ref{fig:LSST_COLORSMEAR}),
the precision is significantly improved for both the
\photoz\ and $\mu$. In the MAIN survey, however,
a redshift-dependent bias remains for $\Zgen > 0.6$.

Although the LSST \photoz\ precision looks promising, 
we urge some caution in the interpretation of these simulations.
If we consider the LSST DEEP-field subset over the same
redshift range as the \SDSS\ sample ($z< 0.4$),
the forecast LSST \photoz\ precision (rms) is about 
$\times 3$ better than that for  the \SDSS\ data sample 
described in \S\ref{sec:results_data}, reflecting the higher 
expected signal to noise and broader wavelength coverage. 
Although we are confident in extrapolating the rms precision
based on the treatment of photon statistics and color smearing,
we cannot rule out unknown systematic effects,
primarily from the unknown source of intrinsic brightness 
variations, that could limit the \photoz\ precision and accuracy.
Concerning the \photoz\ bias presented here, 
this should be considered a lower limit because the same 
light curve model has been used in both the
simulation and in the \zLCFIT\ fit. 
The current \mlcs\ and \SALTII\ models have been shown
to differ significantly in the ultraviolet region (K09),
but such modeling errors have not been considered here.
In future studies, fitting \SALTII\ simulations with \mlcs\ 
(and vice-versa) would likely give an upper limit on the 
\photoz\ bias, and may lead to additional clues about
problems in the light curve models.

\section{Comparisons with Color-Based Redshift Estimates}
\label{sec:compare}

Here we compare our \zLCFIT\ results with color-based
SN redshift estimates for 
the SNLS sample and for an \mlcs-based simulation.

\subsection{Comparison with the SNLS Sample}
\label{sec:compare_SNLS}

\citet{YWang2007_SNLS} determined color-based
photometric redshifts for 40 SNe~Ia from the
SNLS A06 sample.
For the 20 SNe used in the training, the author finds $\zrms = 0.03$;
for the remaining 20 SNe, $\zrms = 0.05$.
In our \zmlcs\ analysis, \NDATASNLSMLCS\  SNe satisfy the
selection criteria, and $\zrms = \zrmsDATASNLSMLCS$.
For the \zSALTII\  method, \NDATASNLSSALT\  SNe satisfy the
selection requirements, and $\zrms = \zrmsDATASNLSSALT$. 
By this metric, 
our \zLCFIT\ method works equally well compared to
the color-based method. Since a list of SNe used by
\citet{YWang2007_SNLS} is not available, we cannot make a more
detailed comparison with the same subset. 

  \subsection{Comparison with \mlcs-based Simulations}
  \label{sec:compare_snrmax25}

We compare \photoz\ results of the two methods on simulations
as described in \citet{YWang2007_SIM} (hereafter WNW07).
Following the procedure in WNW07, 
we simulate observer-frame filters $riz$ using the 
\mlcs\ model for redshifts $\Z < 0.95$ 
(so that $r$-band data are always well defined within the model),
fix the shape-luminosity parameter $\Delta=0$, 
generate a flat redshift distribution,
and ignore intrinsic magnitude variations that introduce 
anomalous Hubble scatter.  
For each SN and each filter, the exposure time is adjusted 
so that ${\rm S/N} = 25$ at the epoch of peak brightness.
In WNW07, only the peak fluxes are used and therefore the
light curve sampling does not matter; 
to investigate the comparative effectiveness of our \zmlcs\ method,
simulated light curves are sampled every 7 days in the 
observer frame.

In the ideal case of no host-galaxy extinction ($A_V=0$), 
WNW07 find  
$\zrms = 0.004$ with a mean bias of $5.4\times 10^{-4}$;
using \zmlcs\ under similar assumptions,
we find $\zrms = \zrmsSIMTESTA$
and a mean bias of $\zbiasSIMTESTA\zbiasUNIT$. 
Including host-galaxy extinction in the simulation  
with a pdf
$P(A_V) = \exp(-A_V/0.46)$ and reddening parameter $R_V=3.1$, 
WNW07 find $\zrms = 0.044$ with a mean bias of $0.008$;
using {\zmlcs}, under the same conditions we find 
$\zrms = \zrmsSIMTESTB$
and a mean bias of $\zbiasSIMTESTB\zbiasUNIT$. 
The significantly improved precision with our \zmlcs\ method 
in these more realistic conditions is in part due to the increase 
in effective signal to noise that comes from using the entire 
light curve.  In addition, shape and color
information contained in the light curve enables
the \zmlcs\ method to partially untangle color variations 
from extinction versus those produced by redshift.

Since the color-based method in WNV07 cannot untangle reddening 
from extinction and redshift, we have attempted a more fair
comparison to the \zmlcs\ method in which 
$A_V$ is fixed to the mean generated
extinction of 0.46~mag; 
in this case $\zrms$ nearly doubles to $\zrmsSIMTESTC$ 
for the \zmlcs\ method,
yet is nearly a factor of 3 smaller than the dispersion in WNV07.
This difference is either due to the enhanced photon statistics
from using the entire light curve in the \zmlcs\ method 
or from non-optimal training in the color-based method.
To increase $\zrms$ to the WNV07 value of 0.044,
the peak S/N must be reduced from 25 to less than 10.

In the \zmlcs\ fits we have used $R_V=3.1$, the same value used 
in generating the simulation. Using the correct value for $R_V$ 
is an apparently unfair advantage over the WNW07 treatment,
in which no assumptions are made about color variations. 
However, in practice the assumption about the value of $R_V$ 
makes little difference to the results: fitting with $R_V=2.2$ 
produces the same precision for $\DZSYM$, although the resulting 
distance moduli are biased by about 0.1~mag.

\begin{figure}[hb]
\centering
 \epsscale{1.}
 \plotone{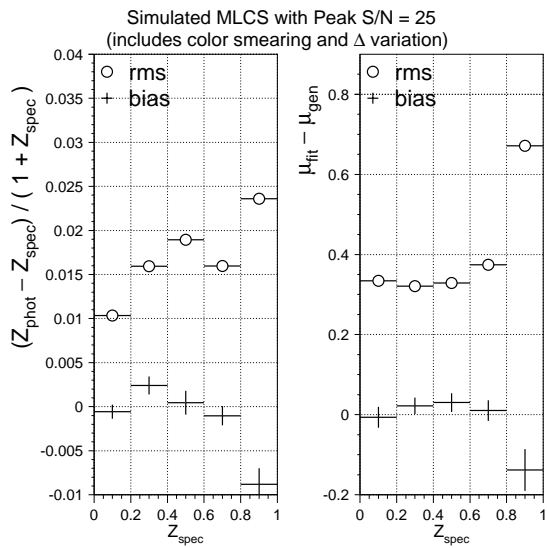}
  \caption{
        For \mlcs\ simulations with peak ${\rm S/N} = 25$,
        fitted \photoz\ bias and $\zrms$ vs. $\Zspec$ (left),
        and distance modulus bias and $\murms$ vs. $\Zspec$ (right).
        The simulations are generated according to the prescription of
        WNW07, but also including 
	non-zero $\Delta$ and intrinsic color variations.
      }
  \label{fig:snrmax25F_bias}
\end{figure}

\begin{figure}[hb]
\centering
 \epsscale{1.1}
 \plotone{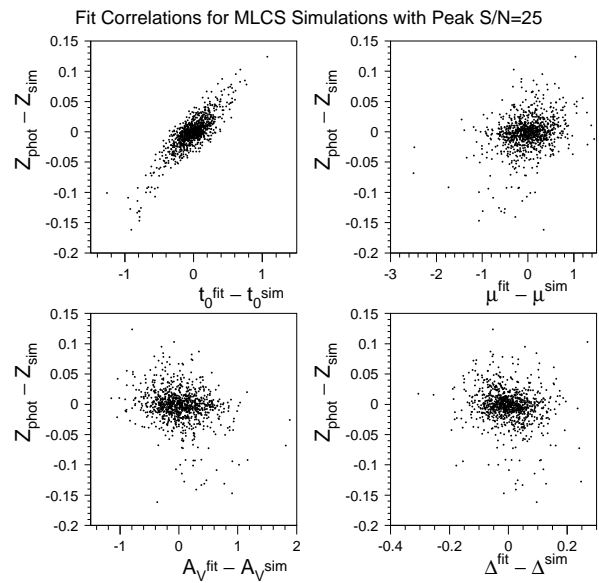}
  \caption{
	Fit-simulation \photoz\ residual ($\Zphot - \Zgen$)
	vs. fit-simulation residual for 
	time of peak brightness ($t_0$),
	distance modulus ($\mu$),
	host-galaxy extinction ($A_V$), and
	shape-luminosity parameter ($\Delta$).
      }
  \label{fig:snrmax25F_corr}
\end{figure}

Even though the effects of host-galaxy extinction are 
included the simulation described in WNW07,
this simulation is not realistic because the shape-luminosity 
parameter $\Delta$ is fixed to zero,
and there are no intrinsic magnitude variations.
Simulating our best estimate for these effects (\S\ref{sec:sim}),
the bias and scatter in the fitted \photoz\ and distance modulus 
are shown as a function of $\Zspec$ in
Fig.~\ref{fig:snrmax25F_bias}.
For redshifts below 0.8
there is no significant bias in either the \photoz\
or distance modulus.
For $\Z > 0.8$ the bias and scatter increase significantly.
The nature of this high-redshift bias is illustrated in
Fig~\ref{fig:snrmax25F_corr}, which shows the correlation between
the fitted versus simulated \photoz\ difference and the fitted
versus simulated difference for
time of peak ($t_0$), extinction, and shape-luminosity parameter.
The \photoz\ outliers are clearly correlated with $t_0$ outliers.
Since the SN~Ia color becomes redder with epoch,
a misestimate of $t_0$ changes the apparent SN color, 
which is translated into an error in the redshift.
The bias in $t_0$ is an artifact of the discrete \mlcs\ passbands
used to characterize the rest-frame light curves.
For SNe simulated at redshifts $\Z > 0.8$, 
the three observer-frame filters map into only 
two rest-frame filters: $riz \to UBB$.
In a small fraction of the light curve fits, 
however, the wrong filter-mapping
($riz \to UBV$) results in a smaller $\chi^2$.
Since the $B$ and $V$-band templates have different shapes,
as well as a 2-day shift in the time of peak brightness, 
the fitted $t_0$ is biased.

\section{Conclusions}
\label{sec:conclude}

We have developed and tested a  photometric redshift estimation method 
using SN Ia light curves
within the framework of the \mlcs\ and \SALTII\ models 
and find that they result in similar \photoz\ precision. We used 
an iterative fitting procedure to determine
the valid observer-frame filters to use in the fits.
Applying this method to \SDSS\ and SNLS data, we obtained
an average \photoz\ precision of $\zrms \sim 0.04$.
To reproduce a comparable level of precision in simulations, 
intrinsic color-smearing is needed  (\S\ref{sec:sim})
at the level of about $0.1$~mag per passband 
or $0.14$~mag per color.
This empirical estimate of color smearing is consistent with 
but does not necessarily imply that there are random color 
variations in SN Ia light curves. 
However, this effect does indicate that there are 
additional sources of color variation that are not 
captured by the \mlcs\ and \SALTII\ light curve models.

We applied the \zLCFIT\ method to simulated LSST samples
(\S\ref{sec:lsst}).
For the DEEP fields, the rms scatter of  $\Zphot - \Zgen$ 
varies from 0.01 to 0.04 without using a host-galaxy \photoz\ prior.
For the MAIN survey the \photoz\ precision is about $\times 2$ worse.
Using a host-galaxy \photoz\ prior significantly reduces outliers and
improves the overall precision.
The next critical step is to apply this method to simulations that
include non-Ia type SNe and estimate the resulting contamination of
photometric SN Ia samples by core-collapse SNe.

\bigskip

We gratefully acknowledge support from the 
Kavli Institute of Cosmological Physics at the University of Chicago,
the National Science Foundation at Wayne State, and the
Department of Energy at Fermilab, the University of Chicago, 
and Rutgers University. S.J. is grateful for the support of DOE 
grant DE-FG02-08ER41562.
Funding for the creation and distribution of the SDSS and SDSS-II
has been provided by the Alfred P. Sloan Foundation,
the Participating Institutions,
the National Science Foundation,
the U.S. Department of Energy,
the National Aeronautics and Space Administration,
the Japanese Monbukagakusho,
the Max Planck Society, and 
the Higher Education Funding Council for England.
The SDSS Web site \hbox{is {\tt \wwwSDSS}.}

The SDSS is managed by the Astrophysical Research Consortium
for the Participating Institutions.  The Participating Institutions are
the American Museum of Natural History,
Astrophysical Institute Potsdam,
University of Basel,
Cambridge University,
Case Western Reserve University,
University of Chicago,
Drexel University,
Fermilab,
the Institute for Advanced Study,
the Japan Participation Group,
Johns Hopkins University,
the Joint Institute for Nuclear Astrophysics,
the Kavli Institute for Particle Astrophysics and Cosmology,
the Korean Scientist Group,
the Chinese Academy of Sciences (LAMOST),
Los Alamos National Laboratory,
the Max-Planck-Institute for Astronomy (MPA),
the Max-Planck-Institute for Astrophysics (MPiA), 
New Mexico State University, 
Ohio State University,
University of Pittsburgh,
University of Portsmouth,
Princeton University,
the United States Naval Observatory,
and the University of Washington.

This work is based in part on observations made at the 
following telescopes.
The Hobby-Eberly Telescope (HET) is a joint project of the 
University of Texas at Austin,
the Pennsylvania State University,  Stanford University,
Ludwig-Maximillians-Universit\"at M\"unchen, and 
Georg-August-Universit\"at G\"ottingen.  
The HET is named in honor of its principal benefactors,
William P. Hobby and Robert E. Eberly.  The Marcario Low-Resolution
Spectrograph is named for Mike Marcario of High Lonesome Optics, who
fabricated several optical elements 
for the instrument but died before its completion;
it is a joint project of the Hobby-Eberly Telescope partnership and the
Instituto de Astronom\'{\i}a de la Universidad Nacional Aut\'onoma de M\'exico.
The Apache Point Observatory 3.5 m telescope is owned and operated by 
the Astrophysical Research Consortium. We thank the observatory 
director, Suzanne Hawley, and site manager, Bruce Gillespie, for 
their support of this project. The Subaru Telescope is operated by the 
National Astronomical Observatory of Japan. The William Herschel 
Telescope is operated by the Isaac Newton Group on the island of 
La Palma in the Spanish Observatorio del Roque 
de los Muchachos of the Instituto de Astrofisica de 
Canarias. The W.M. Keck Observatory is operated as a scientific partnership 
among the California Institute of Technology, the University of 
California, and the National Aeronautics and Space Administration. 
The Observatory was made possible by the generous financial support 
of the W. M. Keck Foundation.

\begin{appendix}

\section{Fitting Issues With Unknown Redshift}
\label{app:prefit}

Here we discuss our implementation of the five pre-fit
issues mentioned in \S\ref{sec:zLCFIT_method}:
(1) SN selection criteria that depend on knowing 
$\Trest$ = $\Tobs/(1+\Z)$
such as requiring measurements with a 
minimum and maximum $\Trest$ value;
(2) determining which observer-frame filters
(with mean wavelength $\lamf$) 
have $\lamf/(1+\Zphot)$ within the
valid wavelength range of the fitting model; 
(3) determining the valid rest-frame epoch range for the 
fitting model;
(4) as $\lamf/(1+\Zphot)$ maps into a different rest-frame filter 
    (for {\mlcs}) there is a discontinuous change in the model error,
    and therefore the $\chi^2$ is not a continuous function of 
    $\Zphot$; and
(5) determining robust initial fit parameter values.

The simplest way to handle SN selection criteria (issue 1) is
to postpone such requirements until the fit has finished
and then use the fitted \photoz\ to determine the $\Trest$ values.
This solution is not practical, however, because we often wish to
remove poorly sampled light curves before fitting, thereby avoiding
pathological fits that are of no interest.
A safe way to apply $\Trest$-dependent requirements before fitting
is to relax such cuts by a factor of $1+\Zmax$, 
where $\Zmax$ is a safe upper bound on all redshifts.
For example, consider the requirements of a measurement
with $\Trest < -6$~days and $21 < \Trest < 60$~days.
Using $\Zmax = 0.5$ for the \SDSS,
the pre-fitted requirements are $\Trest < -4$~days
and $14 < \Trest < 90$~days.
Any initial redshift value can be used to 
determine $\Trest$ as long as it is less than $\Zmax$.
Although these $\Trest$-related requirements are relaxed,
they are still useful for rejecting poorly sampled
light curves; the nominal cut is applied after the fit
using the fitted \photoz\ value.

To determine the valid observer-frame filters (issue 2),
the first-iteration \photoz\ fit uses all filters,
with a possible exception for those covering the 
ultraviolet region with a mean wavelength below about 4000~\AA. 
The ultraviolet filter should be left out of the first iteration
if it is rarely used, noting that it can be added back
for the second fit iteration. The fundamental assumption
about the fitting model is that extrapolating beyond the
defined wavelength range gives reasonable magnitude estimates
so that the fit converges and that the \photoz\
bias is not too large.
The (biased) \photoz\ estimate
is then used to determine which observer-frame filters to
retain and to reject for the second fit iteration.
Note that if an ultraviolet filter is excluded initially, 
then a low \photoz\ value will result in the inclusion
of this filter for the second iteration.
If a filter is excluded, the $\Trest$-related requirements
are re-tested; the fit stops if the light curve no longer has 
adequate sampling.

Since the first-iteration \photoz\ ($\Zphotone$)
is biased and has some \unc, a safety margin is used
to determine which observer-frame filters to keep.
For each observer-frame filter (f) with mean wavelength $\lamf$, 
the valid redshift range for this filter is defined by 
\begin{equation}
  \zminf =  \lamf/\lammax^{\rm model} - 1  ~~~~~~~~~~~~~~~~
  \zmaxf =  \lamf/\lammin^{\rm model} - 1  ~,
  \label{eq:zmaxf}
\end{equation}
where $\lambda^{\rm model}_{\rm min,max}$ are the minimum and maximum
rest-frame wavelengths defined by the model.
For example, the \SALTII\ model is defined for rest-frame wavelengths
2900--7000~\AA; the observer-frame $i$-band 
($\bar{\lambda}_i = 7500$~{\AA}) 
is therefore valid for redshifts above $\zmini = 0.071$,
and $g$-band ($\bar{\lambda}_g = 4720~{\AA}$)  is valid
for redshifts below $\zmaxg = 0.63$.
For this analysis, a filter is kept in the second iteration
if the first-iteration fitted \photoz\ satisfies
\begin{equation}
   \zminf + \cutZphotSYM < \Zphotone < \zmaxf - \cutZphotSYM ~.
   \label{eq:Zphotone}
\end{equation}
We have set $\cutZphotSYM = \cutphotodzSNLS$ and $\cutphotodzSDSS$
for the SNLS and SDSS surveys, respectively.
$\cutZphotSYM$ is smaller for the {\SDSS} because the
S/N for SNe at $\zmini = 0.07$ is larger than the S/N
for SNLS SNe at $\zmaxg = 0.063$.
The $\cutZphotSYM$ filter-selection cut has not been optimized;
the optimal cut is likely to depend on the \unc\ of the host-galaxy
\photoz\ and may also depend on redshift.

\begin{figure}  
\centering
 \epsscale{.8}
 \plotone{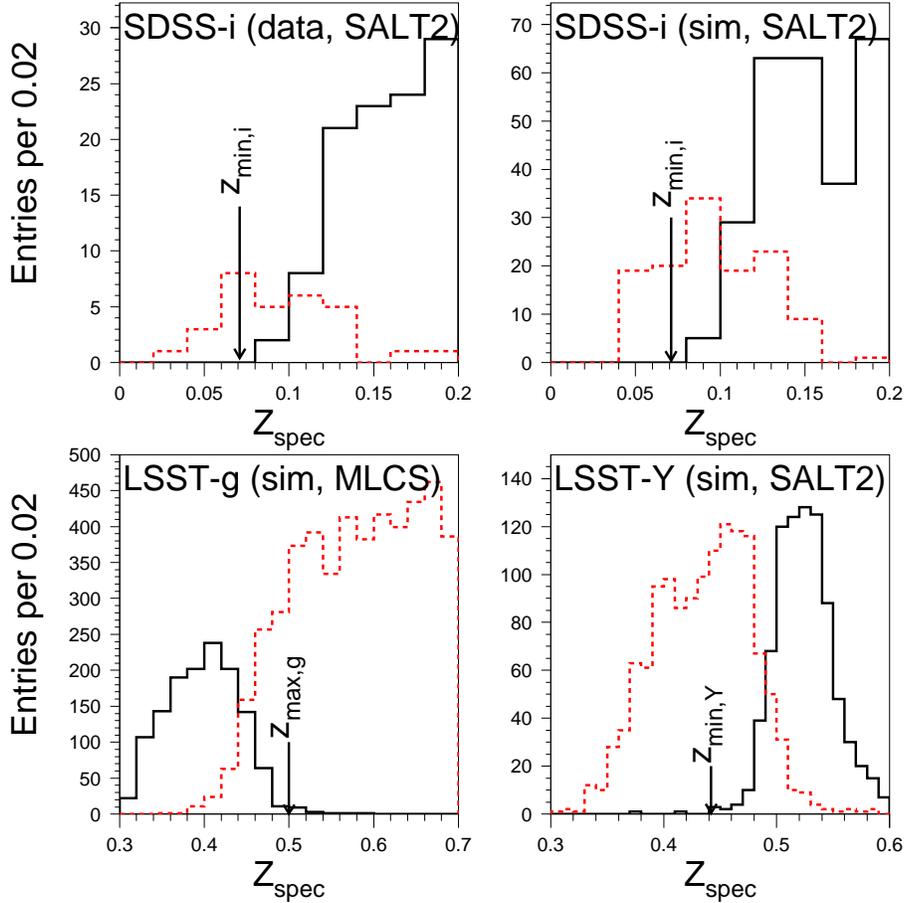}
  \caption{
	Illustration of filters excluded after the 
	first fit iteration. Each plot shows the \spec\ 
	(i.e., true) redshift distribution when the 
	indicated filter is kept in the \photoz\ fit (solid) 
	and when the same filter has been 
	dropped after the first fit iteration (dashed). 
	The vertical arrow shows either $\zminf$ or $\zmaxf$ for 
	the filter indicated. 
	$\Zspec$ is used only in making these plots, and is not
	used in the \photoz\ fits.
      }
  \label{fig:zdrop}
\end{figure}

This filter selection algorithm is illustrated in Fig.~\ref{fig:zdrop}.
The solid histograms in the top two plots (SDSS data and simulation) 
show the true redshift ($\Zspec$) distributions near $\zmini$ when 
the $i$-band is kept in the \zSALTII\ fit. 
When the $i$-band is discarded (dashed), the entire light curve
is discarded because of the requirement of having three filters.
The gap between $\zmini = 0.071$ and the bulk
distribution is due to the $\cutZphotSYM$ cut.
The bottom two plots in Fig.~\ref{fig:zdrop} illustrate
cases for LSST when a dropped filter does not result in 
rejecting the entire light curve.
When the LSST $g$-band is included in the \zmlcs\ fit 
(solid curve, lower left), 
the true redshift almost always satisfies the model-validity
requirement ($\Zspec < 0.5$).
The dashed curve shows that the $g$-band has been correctly
excluded when the true redshift is above 0.5,
but we have also excluded this passband in some 
cases where it is valid (i.e., $\Zspec < 0.5$).
The situation is similar for the LSST $Y$-band (lower right plot).
For both passbands, a small number of light curve fits include
these bands when they should have been excluded;
potential biases from these invalid passbands should be
accounted for in the assessment of 
systematic \uncs\ on cosmological parameters.

The selection of rest-frame epochs (issue 3), 
which depends on $\Tobs/(1+\Zphot)$,
is made in the same way as the selection of filters (issue 2).
All measurements are included in the first fit iteration,
and the initial \photoz\ estimate is then used to select
which epochs satisfy the epoch range of the model.

The $\chi^2$ continuity (issue 4) is an issue because the
\snana\ minimization  is based on {\minuit},
and this program relies on computing local derivatives 
with respect to the fitting parameters. This means that
discontinuities in the $\chi^2$ function and its derivative can
lead to fitting failures or pathological results.
The problem can be solved by marginalizing, 
but the necessary five-dimensional integration requires 
$\sim 100$ times more CPU processing. 
In addition to the utility of using the much faster minimization program, 
it is useful to identify $\chi^2$ discontinuities because such 
functional pathologies usually correspond to unphysical behavior 
in the light curve model, and the model should be improved accordingly.

There are two main sources of $\chi^2$ discontinuity:
(1) interpolating lookup tables, and 
(2) filter-dependent model parameters.
For class (1), appropriate numerical methods must be used to ensure
continuity in both the function and its derivative.
For class (2), the problem occurs for rest-frame models such as
{\mlcs}, 
in which an infinitesimal change in $\Zphot$ results in a different
rest-frame filter for the model or a different color to use for
spectral warping in the $K$-corrections.
This step-function change in the model parameters results
in a discontinuity in the model magnitude (or its error) as
a function of $\Zphot$. To prevent such sharp discontinuities,
\snana\ uses a smooth transition function 
($A+B\tan^{-1}({\lambda}/{\tau}_{\lambda})$) 
to smoothly vary model parameters between 
neighboring filters.

Since the redshift and SN color are somewhat degenerate,
initial parameter estimation (issue 5) 
is important so that \minuit\ will find a global, 
rather than local, minimum.  
The initial color and \photoz\ are determined from a crude 
$\chi^2$ minimization on a coarse grid:
the color is varied in bins of $\CBINSCAN$  and the \photoz\
is varied in bins of $\ZBINSCAN$.
For each \photoz\ value, an initial distance modulus estimate
($\muini$) is needed to ensure robust convergence of {\sc minuit}.
Although $\muini$ is calculated from a standard $\Lambda$CDM cosmology
using a specific set of cosmological parameters, 
the fitted distance modulus is unconstrained and is therefore
not biased by the $\muini$ calculation.
For \SALTII, the initial $x_0$ value is estimated by 
inverting Eq.~\ref{eq:muSALTII}.
The shape-luminosity parameter is initialized to an average value: 
$\Delta = -0.1$ for \mlcs, and $x_1=0.0$ for \SALTII.
After the first fit iteration, the fitted parameters are used
as initial values for the second fit iteration.
When a filter is dropped and the first fit iteration is repeated,
the fitted parameters are used as initial values, even though
an invalid filter was used in the fit. 
If we do not use a coarse grid to initialize the color and \photoz,
and simply start with average values, the \photoz\ precision
for the \SDSS\ sample is only slightly degraded.
For the SNLS sample, however, which has a much larger redshift
range compared to the \SDSS, the \photoz\ precision is degraded 
by a factor of 2 due to a significant number of catastrophic 
outliers.

\section{Marginalization}
\label{app:marg}

Here we describe some of the details related to 
the \marg.
The minimized values and \uncs\ are used to estimate 
the integration ranges: $\pm 4\sigma$ around
the minimized value for each parameter.
The integration grid consists of $\Ngrid$
points per fit-parameter or a total of 
${\Ngrid}^5$ integration cells. We find
that $\Ngrid=11$ is a good compromise between
precision and computing time.
To improve calculation speed per integration cell,
the $\chi^2$ calculation stops when the probability 
falls below $10^{-5}$.

The \marg\ is repeated for either of the following cases:
(1) the probability at any integration boundary
is greater than 0.03, or 
(2) more than three one-dimensional bins 
(i.e., marginalized over the other four parameters)
have a probability less than $10^{-4}$.
In the first case the integration range is extended,
while in the second case the integration range is
reduced.

\end{appendix}


\vspace{1cm}
\bibliographystyle{apj}
\bibliography{lcfitz}

\begin{thebibliography}{31}
\expandafter\ifx\csname natexlab\endcsname\relax\def\natexlab#1{#1}\fi

\bibitem[{Abdalla {et~al.}(2008)}]{Abdalla08}
Abdalla, F. {et~al.} 2008, submitted, arXiv:0812.3831

\bibitem[{Astier {et~al.}(2006)}]{Astier06}
Astier, P. {et~al.} 2006, \aap, 447, 31

\bibitem[{Benitez(2000)}]{Benitez2000}
Benitez, N. 2000, \apj, 536, 571

\bibitem[{{Bernstein} {et~al.}(2009){Bernstein}, {Kessler}, {Kuhlmann}, \&
  {Spinka}}]{DES-moriond2009}
{Bernstein}, J.~P., {Kessler}, R., {Kuhlmann}, S., \& {Spinka}, H. 2009,
  arXiv:0906.2955

\bibitem[{Dahlen \& Goobar(2002)}]{Dahlen2002}
Dahlen, T. \& Goobar, A. 2002, \pasp, 114, 284

\bibitem[{{Delgado} {et~al.}(2006){Delgado}, {Cook}, {Miller}, {Allsman}, \&
  {Pierfederici}}]{OPSIM2006}
{Delgado}, F., {Cook}, K., {Miller}, M., {Allsman}, R., \& {Pierfederici}, F.
  2006, in Proc. SPIE, Vol. 6270

\bibitem[{Frieman {et~al.}(2008)}]{Frieman07}
Frieman, J.~A. {et~al.} 2008, \aj, 135, 338

\bibitem[{{Fukugita} {et~al.}(1996)}]{Fukugita_96}
{Fukugita}, M. {et~al.} 1996, \aj, 111, 1748

\bibitem[{Gong {et~al.}(2010)Gong, Cooray, \& Chen}]{Gong09}
Gong, Y., Cooray, A., \& Chen, X. 2010, \aj, 709, 1420

\bibitem[{Gunn {et~al.}(1998)}]{Gunn_98}
Gunn, J.~E. {et~al.} 1998, \aj, 116, 3040

\bibitem[{Gunn {et~al.}(2006)}]{SDSS_telescope}
---. 2006, \aj, 131, 2332

\bibitem[{Guy {et~al.}(2007)}]{Guy07}
Guy, J. {et~al.} 2007, \aap, 466, 11

\bibitem[{Holtzman {et~al.}(2008)}]{Holtz08}
Holtzman, J. {et~al.} 2008, \aj, 136, 2306

\bibitem[{Hsiao {et~al.}(2007)}]{Hsiao07}
Hsiao, E. {et~al.} 2007, \apj, 663, 1187

\bibitem[{{Ivezi{\'c}} {et~al.}(2008)}]{Ivezic_08}
{Ivezi{\'c}}, {\v Z}. {et~al.} 2008, arXiv:0805.2366

\bibitem[{{Jha} {et~al.}(2007){Jha}, {Riess}, \& {Kirshner}}]{JRK07}
{Jha}, S., {Riess}, A.~G., \& {Kirshner}, R.~P. 2007, \aj, 659, 122

\bibitem[{Kessler {et~al.}(2009{\natexlab{a}})}]{K09}
Kessler, R. {et~al.} 2009{\natexlab{a}}, \apjs, 185, 32

\bibitem[{Kessler {et~al.}(2009{\natexlab{b}})}]{SNANA09}
---. 2009{\natexlab{b}}, \pasp, 121, 1028

\bibitem[{Kim \& Miquel(2007)}]{KimMiquel07}
Kim, A. \& Miquel, R. 2007, Astropart. Phys., 28, 448

\bibitem[{LSST~Science~Collaborations(2009)}]{LSSTSB09}
LSST~Science~Collaborations, A. 2009, in {arXiv:0912.0201}

\bibitem[{Lupton {et~al.}(2001)}]{Lupton_01}
Lupton, R. {et~al.} 2001, in ASP Conf. Ser. 238: Astronomical Data Analysis
  Software and Systems X, ed. F.~R. {Harnden}, Jr., F.~A. {Primini}, \& H.~E.
  {Payne}, 269--+

\bibitem[{Nobili \& Goobar(2008)}]{Nobili2007}
Nobili, S. \& Goobar, A. 2008, \aap, 487, 19

\bibitem[{Nugent {et~al.}(2002)}]{Nugent2002}
Nugent, P. {et~al.} 2002, \pasp, 114, 803

\bibitem[{Palanque-Delabrouille {et~al.}(2009)}]{SNLS_PHOTOZ09}
Palanque-Delabrouille, N. {et~al.} 2009, arXiv:0911.1629

\bibitem[{Sako {et~al.}(2008)}]{Sako08}
Sako, M. {et~al.} 2008, \aj, 135, 348

\bibitem[{Schlegel {et~al.}(2009)}]{BOSS_09}
Schlegel, D. {et~al.} 2009, arXiv:0902.4680

\bibitem[{Sullivan {et~al.}(2006)}]{Sullivan2006}
Sullivan, M. {et~al.} 2006, \aj, 131, 969

\bibitem[{Wang(2007)}]{YWang2007_SNLS}
Wang, Y. 2007, \apj, 654, L123

\bibitem[{Wang {et~al.}(2007)Wang, Narayan, \& Wood-Vasey}]{YWang2007_SIM}
Wang, Y., Narayan, G., \& Wood-Vasey, M. 2007, \mnras, 382, 377

\bibitem[{{York} {et~al.}(2000)}]{York_00}
{York}, D.~G. {et~al.} 2000, \aj, 120, 1579

\bibitem[{Zheng {et~al.}(2008)}]{Zheng08}
Zheng, C. {et~al.} 2008, \aj, 135, 1766

\end{thebibliography}

  \end{document}